\documentclass[journal]{IEEEtran}
\usepackage[bookmarks=true,linkcolor=blue,citecolor=blue,urlcolor=blue]{hyperref}
\usepackage{cite}
\usepackage{amsmath,amssymb,amsfonts}
\usepackage{amsthm}
\usepackage{algorithmic}
\usepackage{algorithm}
\usepackage{graphicx}
\usepackage[caption=false]{subfig}
\usepackage{textcomp}
\usepackage{float}
\usepackage{booktabs}
\usepackage{basicNotation}
\usepackage{mathtools}
\usepackage{bm}

\providecommand{\transpose}{^{\mathsf{T}}}

\providecommand{\diag}{\mathrm{diag}}

\newcommand{\fk}{~\forall~k \in \cK}
\newcommand{\cQ}{\mathcal Q}

\begin{document}

\title{A Duality-Based Fixed Point Iteration Algorithm for Transmit Beamforming Design in ISAC Systems}

\author{\IEEEauthorblockN{Xilai~Fan and Ya-Feng~Liu}
\thanks{X. Fan is with the Academy of Mathematics and Systems Science, Chinese Academy of Sciences, Beijing 100190, China, and also with the School of Mathematical Sciences, University of Chinese Academy of Sciences, Beijing, China (e-mail: fanxilai@lsec.cc.ac.cn).}
\thanks{Y.-F. Liu is with the Ministry of Education Key Laboratory of Mathematics and Information Networks, School of Mathematical Sciences, Beijing University of Posts and Telecommunications, Beijing 102206, China (e-mail: yafengliu@bupt.edu.cn).}
}

\maketitle


\begin{abstract}
This paper considers the transmit beamforming design problem in an integrated sensing and communication (ISAC) system, where a multi-antenna base station simultaneously serves multiple communication users and performs radar sensing.
The problem is formulated as the minimization of the total transmit power subject to signal-to-interference-plus-noise ratio (SINR) constraints for communication users and mean-squared-error (MSE) constraints for radar sensing.
We show that the semidefinite relaxation (SDR) of the original problem is tight, derive its Lagrangian dual, and reformulate it as a generalized downlink beamforming (GDB) problem with potentially indefinite weighting matrices.
Compared to the classical downlink beamforming problem, the indefinite weighting matrices in the GDB problem cause the problem to be possibly unbounded and prevent the direct application of classical fixed point iteration (FPI) algorithms whose convergence relies on the standard interference function theory.
To address these challenges, we first derive a necessary and sufficient condition for the boundedness of the GDB problem and then develop a tailored FPI algorithm with a comprehensive theoretical analysis, including global convergence guarantees, stability characterization of the fixed points, and a linear convergence rate. 
Since the FPI algorithm requires appropriate initial points, we further propose three initialization strategies that offer different trade-offs between theoretical guarantees
and computational costs.
Building upon all of previous results, we propose a duality-based FPI (Dual-FPI) algorithm that integrates an outer subgradient ascent loop with an inner FPI loop, and prove its overall convergence. 
Numerical results demonstrate that the proposed Dual-FPI algorithm achieves globally optimal solutions while being about two orders of magnitude faster than existing semidefinite programming based approaches.
\end{abstract}


\section{Introduction}\label{sec:introduction}

\IEEEPARstart{I}{ntegrated} sensing and communications (ISAC) has emerged as a key enabling technology for 6G wireless networks, facilitating the simultaneous realization of communication and sensing functionalities using shared hardware and spectrum resources \cite{liutcom202,liujsac2022,chiriyath2017radar,zhang2021overview,cui2021integrating,liu2024SurveyRecentAdvances}.
Among various ISAC design tasks, transmit beamforming is widely regarded as a core problem, as it directly determines both the spatial distribution of transmitted energy and the interference patterns experienced by communication users and sensing targets \cite{liu2018toward,liu2018mumimo,liu2020JointTransmitBeamforming,liu2022TransmitDesignJoint,liu2022CramerRaoBoundOptimization,wen2023EfficientTransceiverDesign,wang2023QoSawarePrecoderOptimization,attiah_beamforming_2024,attiah2025uplink-downlink,zhang_joint_2025}.

In conventional communication-only multi-user downlink systems, the classical downlink beamforming (DB) problem, which minimizes the total transmit power subject to signal-to-interference-plus-noise ratio (SINR) constraints, can be efficiently solved using uplink--downlink duality \cite{rashidDL1998,boche2002GeneralDualityTheory,wiesel2006LinearPrecodingConic}.
This duality establishes that the minimum sum power required to achieve a set of SINR targets in the downlink is equal to that in a virtual dual uplink channel, and the resulting virtual uplink problem can be solved globally and efficiently via fixed point iteration (FPI) algorithms.

However, beamforming problems in ISAC systems exhibit significant structural differences from those in communication-only systems.
Radar sensing requirements are often expressed via beampattern matching \cite{li2007mimo,stoica2007probing,liu2020JointTransmitBeamforming,liu2022TransmitDesignJoint} or Cram\'{e}r--Rao bound (CRB) conditions \cite{liu2022CramerRaoBoundOptimization,attiah_beamforming_2024}, which are nonlinear functions of the transmit covariance matrix.
These constraints couple the beamforming vectors across all communication and sensing signals, resulting in optimization problems in which the associated dual weighting matrices can be indefinite \cite{attiah_beamforming_2024,attiah2025uplink-downlink,zhang_joint_2025}.
The presence of indefinite weighting matrices introduces two fundamental technical difficulties.
First, the optimization problem might become unbounded.
Second, even when the problem is bounded, the classical FPI algorithm cannot be directly applied, because its convergence analysis relies on the standard interference function theory \cite{yates1995FrameworkUplinkPower}, which requires positive definite weighting matrices.
The goal of this paper is to overcome these difficulties and extend efficient duality-based beamforming approaches from communication-only systems to ISAC systems.

\subsection{Prior Works}

In communication-only systems, the uplink--downlink duality for the DB problem, relating the downlink SINR region to that of a virtual uplink channel, was established in \cite{rashidDL1998,visotsky1999OptimumBeamformingUsing,rashid-farrokhi1998JointOptimalPower,boche2002GeneralDualityTheory,Viswanathsum2003,Vishwanathduality2003,yuminimax2006,song2007NetworkDualityMultiuser}.
Building on this duality, \cite{Schubertsolution2004,schubert2005IterativeMultiuserUplink} developed globally optimal alternating optimization algorithms that solve a power allocation subproblem exactly, and \cite{wiesel2006LinearPrecodingConic} improved computational efficiency by replacing this step with FPI updates.
Using the nonlinear Perron--Frobenius theory \cite{krause1986PerronStabilityTheorem,lemmens2012NonlinearPerronFrobeniusTheory}, the work \cite{cai2011MaxminWeightedSINR} proved that the FPI algorithm in \cite{wiesel2006LinearPrecodingConic} is guaranteed to find the global solution.
These algorithms fall within the standard interference function framework \cite{yates1995FrameworkUplinkPower}, which provides a unified convergence theory for FPI-based power control.
The duality framework has also been extended to per-antenna power constraints \cite{yu2007TransmitterOptimizationMultiantenna,fan2025adaptive}, indefinite shaping constraints \cite{hammarwall2006DownlinkBeamformingIndefinite}, and coordinated multicell systems \cite{dahrouj2010CoordinatedBeamformingMulticell}.
More recently, the Lagrangian duality approach has been extended to cooperative cellular networks with rate-limited fronthaul links \cite{fan2022EfficientlyGloballySolving,liu2021UplinkdownlinkDualityMultipleaccess,fan2025qos}, where the FPI algorithm was shown to achieve global optimality with a linear convergence rate \cite{fan2025qos}.

In the context of ISAC beamforming design, the pioneering works \cite{liu2018toward,liu2018mumimo} investigated dual-functional radar-communication (DFRC) waveform design, where the transmit signal is jointly optimized for communication and radar functionalities.
For MIMO radar beampattern design, the works \cite{li2007mimo,stoica2007probing} developed signal processing techniques that form the sensing performance metrics adopted by subsequent ISAC studies.
Building on these foundations, existing ISAC beamforming studies have primarily explored two approaches.
One line of research employs the semidefinite relaxation (SDR) technique \cite{luo2010SemidefiniteRelaxationQuadratic,bengtsson2002OptimumSuboptimumTransmit} to transform the original nonconvex problem into a tractable form.
The tightness of the SDR has been established under various system settings \cite{liu2020JointTransmitBeamforming,liu2022TransmitDesignJoint,liu2022CramerRaoBoundOptimization,wang2023QoSawarePrecoderOptimization,huang2010rank,zhang_joint_2025}, but SDR-based approaches often incur high computational costs due to the dimension lifting involved.
Another line of research has extended the uplink--downlink duality to ISAC systems \cite{attiah_beamforming_2024,attiah2025uplink-downlink,zhang_joint_2025,zhang2024optimal,zhu2024joint}.
The works \cite{attiah_beamforming_2024,attiah2025uplink-downlink} derived a generalized duality result under the assumption of no dedicated sensing beamformer, and characterized the admissibility conditions under which the resulting generalized downlink beamforming (GDB) problem is bounded.
The works \cite{zhang_joint_2025,zhang2024optimal} extended this duality framework to networked ISAC systems with fronthaul-rate constraints, where the sensing requirement is captured by a scalar SINR constraint and yields a scalar sensing dual variable.
A related work \cite{zhu2024joint} considered networked sensing without communication constraints, exploiting the resulting decomposable structure between beamforming and fronthaul compression.
However, none of these works provides an efficient FPI algorithm with provable convergence guarantees, analogous to those available for communication-only systems, for general ISAC beamforming problems with indefinite weighting matrices.

\subsection{Our Contributions}

In this paper, we study the transmit power minimization problem subject to SINR constraints for communication users and mean-squared-error (MSE) constraints for radar sensing.
By leveraging SDR and Lagrangian duality, we reformulate the original ISAC beamforming problem into a GDB problem with potentially indefinite weighting matrices, and develop efficient algorithms with provable convergence guarantees.
The main contributions of this paper are as follows.

\begin{itemize}
	\item \emph{Structural Analysis of the GDB Problem.}
	We derive the Lagrangian dual of the GDB problem and establish its key structural properties.
	In particular, we provide a necessary and sufficient condition for the boundedness of the GDB problem, which resolves the feasibility issues caused by indefinite weighting matrices and provides a theoretical foundation for algorithmic design.

	\item \emph{FPI Algorithm with Convergence and Stability Guarantees.}
	We develop an FPI algorithm tailored to the GDB problem and provide a comprehensive theoretical analysis.
	More specifically, we prove that the proposed FPI converges globally to the optimal solution under a mild initialization condition.
	We further characterize the stability of the fixed points, showing that the maximal fixed point, which corresponds to the optimal dual solution, is the unique stable fixed point.
	We also establish the linear convergence rate of the proposed FPI.
	Building on these results, we prove the overall convergence of the Dual-FPI algorithm.

	\item \emph{Initialization Strategies.}
	We propose three initialization strategies, the SDP-based method, the warm-start method, and the heuristic method, with different trade-offs between theoretical guarantees and computational costs.
	In particular, the SDP-based method provides a rigorous feasibility guarantee, the warm-start method exploits continuity across outer iterations, and the heuristic method offers a computationally cheap alternative based on the single-user analysis.
\end{itemize}

In our prior work \cite{fan2026duality}, we presented a preliminary version of the duality-based FPI algorithm for the ISAC beamforming problem.
The present paper is a significant extension of \cite{fan2026duality}.
First, we provide complete convergence and stability analyses of the fixed points determined by the fixed point equation, including the characterization of multiple fixed points and the identification of the unique stable fixed point.
Second, we establish the linear convergence rate of the FPI and the overall convergence of the Dual-FPI algorithm.
Third, we develop three initialization strategies and empirically compare their performance.
Fourth, we conduct extensive numerical experiments comparing the proposed algorithm with state-of-the-art benchmarks.

\subsection{Notation and Organization}

We adopt the following notation throughout this paper.
We use lowercase boldface letters for vectors and uppercase boldface letters for matrices.
The symbols $(\cdot)^\hermitian$, $(\cdot)^\transpose$, and $(\cdot)^{-1}$ denote the conjugate transpose, transpose, and inverse, respectively.
We use $\tr(\cdot)$ to denote the trace of a matrix and $\diag(\cdot)$ to denote a diagonal matrix.
For two Hermitian matrices $\m A$ and $\m B$, $\m A \succeq \m B$ and $\m A \succ \m B$ mean that $\m A - \m B$ is positive semidefinite and positive definite, respectively.
We use $\m I$ to denote the identity matrix of an appropriate size and $\rho(\cdot)$ to denote the spectral radius.
The order relationship between two vectors is understood component-wise.

The remainder of this paper is organized as follows.
Section~\ref{sec:system_model} presents the system model and problem formulation.
Section~\ref{sec:dual_algorithm} develops the duality-based algorithm framework, including the SDP reformulation and the Lagrangian dual analysis.
Section~\ref{sec:inner} proposes the FPI algorithm for solving the GDB problem and provides the convergence, stability, and convergence rate analyses.
Section~\ref{sec:initialization} presents the initialization strategies.
Section~\ref{sec:numerical} reports numerical experiments.
Section~\ref{sec:conclusion} concludes the paper.

\section{System Model and Problem Formulation}\label{sec:system_model}

Consider a multi-user ISAC system where a base station~(BS) equipped with $M$ antennas serves $K$ single-antenna communication users while simultaneously performing radar sensing. Let $\cK = \{1, 2, \dots, K\}$ denote the set of communication users. The transmitted signal is given by
\begin{equation}\label{eq:signal}
\v x = \begin{bmatrix}
	\v v_1, \v v_2, \d, \v v_K
\end{bmatrix} \v s + \m V_s \tilde{\v s},
\end{equation}
where $\v s \in \mathbb{C}^K$ is the communication signal, $\tilde{\v s} \in \mathbb{C}^M$ is the sensing signal, $\v v_k \in \mathbb{C}^{M}$ is the beamforming vector for user $k$, and $\m V_s \in \mathbb{C}^{M \times M}$ represents the beamforming matrix for sensing \cite{liu2020JointTransmitBeamforming, liu2022CramerRaoBoundOptimization, liu2022TransmitDesignJoint}.
Under the signal model in \eqref{eq:signal}, the transmit covariance matrix $\m R$, which captures the second-order statistics of the transmitted signal, can be expressed as
\begin{equation}
\label{eq:R}
\m R = \sum_{k \in \cK} \v v_k \v v_k^\hermitian + \m V_s \m V_s^\hermitian.
\end{equation}

\subsection{Communication Model}
The received signal at communication user $k$ is given by
\begin{equation*}
y_k = \v h_k^\hermitian \v x + n_k = \v h_k^\hermitian \v v_k s_k + \sum_{j \neq k} \v h_k^\hermitian \v v_j s_j + \v h_k^\hermitian \m V_s \tilde{\v s} + n_k,
\end{equation*}
where $\v h_k \in \mathbb{C}^{M}$ is the channel vector from the BS to user $k$, and $n_k \sim \mathcal{CN}(0, \sigma_k^2)$ is the additive noise. The SINR for user $k$ can be expressed as
\begin{equation}
\text{SINR}_k = \frac{|\v h_k^\hermitian \v v_k|^2}{\sum_{j \neq k} |\v h_k^\hermitian \v v_j|^2 + \v h_k^\hermitian \m V_s \m V_s^\hermitian \v h_k + \sigma_k^2}.\label{eq:SINR}
\end{equation}

\subsection{Sensing Model}
For radar sensing performance, we employ the MSE between the designed and desired beampatterns as the performance metric \cite{li2007mimo,stoica2007probing,liu2020JointTransmitBeamforming}.
The MSE is defined as
\begin{equation}
E(\alpha, \m R) = \frac{1}{Q} \sum_{q \in \cQ} \left| \alpha d(\theta_q) - \mathbf{a}(\theta_q)^\hermitian \m R \mathbf{a}(\theta_q) \right|^2,
\label{eq:mse}
\end{equation}
where $\alpha \geq 0$ is a scaling factor, $\m R$ is defined in \eqref{eq:R}, $\mathcal Q = \{1, 2, \d, Q\}$, $\{\theta_q\}_{q \in \cQ}$ represents the sampled angle grid, $\mathbf{a}(\theta) \in \mathbb{C}^{M}$ is the steering vector at angle $\theta$, and $d(\theta) \in \mathbb{R}$ denotes the desired beampattern.
For a given transmit covariance matrix $\m R$, the optimal scaling factor $\alpha^\star$ that minimizes the MSE can be obtained in closed form as
\begin{equation}
\alpha^\star = \frac{\sum_{q \in \cQ} d(\theta_q) \left( \mathbf{a}(\theta_q)^\hermitian \m R \mathbf{a}(\theta_q) \right)}{\sum_{q \in \cQ} d^2(\theta_q)}.
\label{eq:alpha}
\end{equation}
By substituting $\alpha^\star$ in \eqref{eq:alpha} back into \eqref{eq:mse}, the MSE is reformulated as a quadratic form in $\m R$:
\begin{equation}
E^\star(\m R) = \frac{1}{Q}\sum_{q \in \cQ} \langle \m R, \m M_q \rangle^2,
\label{eq:mse_const}
\end{equation}
where the matrices $\{\m M_q\}$ are given by
\begin{equation*}
\m M_q = \frac{d(\theta_q) \sum_{q' \in \cQ} d(\theta_{q'}) \mathbf{a}(\theta_{q'}) \mathbf{a}(\theta_{q'})^\hermitian}{\sum_{q' \in \cQ} d^2(\theta_q')} - \mathbf{a}(\theta_q) \mathbf{a}(\theta_q)^\hermitian.
\end{equation*}

\subsection{Problem Formulation}

The objective is to minimize the total transmit power while guaranteeing communication SINR constraints and radar sensing performance.
The ISAC beamforming problem can be formulated as:
\begin{equation}
\begin{aligned}
\min_{\{\v v_k\}, \m V_s} \quad & \sum_{k \in \cK} \|\v v_k\|^2 + \tr (\m V_s \m V_s^\hermitian) \\
\st~~\quad & \text{SINR}_k \geq \gamma_k, \fk, \\
& E^\star(\m R) \leq \eta,
\end{aligned}
\label{eq:jsc}
\end{equation}
where $\{\gamma_k > 0\}$ is the SINR target of user $k$ that guarantees the communication requirement, and $\eta > 0$ is the MSE constraint that ensures radar sensing performance.

By substituting the SINR expression in \eqref{eq:SINR} and MSE expression in \eqref{eq:mse_const}, problem \eqref{eq:jsc} can be explicitly written as:
\begin{equation}
\begin{aligned}
\min_{\{\v v_k\}, \m V_s} \quad & \sum_{k \in \cK} \|\v v_k\|^2 + \tr (\m V_s \m V_s^\hermitian) \\
\st~~\quad & \left(1 + \frac{1}{\gamma_k}\right) |\v h_k^\hermitian \v v_k|^2 \geq \sum_{j \in \cK} |\v h_k^\hermitian \v v_j|^2 + \v h_k^\hermitian \m V_s \m V_s^\hermitian \v h_k \\
&\quad\quad\quad\quad\quad\quad\quad\quad+ \sigma_k^2, \fk, \\
& \sum_{q \in \cQ} \left\langle \sum_{k \in \cK} \v v_k \v v_k^\hermitian + \m V_s \m V_s^\hermitian, \m M_q \right\rangle^2 \leq Q \eta.
\end{aligned}
\label{eq:jsc_exp}
\end{equation}
Problem \eqref{eq:jsc_exp} is nonconvex due to the quadratic SINR constraints and the quartic sensing constraint, making it difficult to solve directly.
To address this, we develop a duality-based algorithm in Section~\ref{sec:dual_algorithm} that decomposes the problem into a sequence of tractable subproblems.
Although this paper focuses on formulation \eqref{eq:jsc_exp}, the developed results can be readily extended to other ISAC beamforming design formulations~(e.g.,~\cite{liu2020JointTransmitBeamforming,liu2022CramerRaoBoundOptimization,attiah_beamforming_2024}).

\section{A Duality-Based Algorithm for Solving Problem \eqref{eq:jsc_exp}}\label{sec:dual_algorithm}

In this section, we develop a duality-based algorithm for solving problem \eqref{eq:jsc_exp}.
The idea is to first reformulate the original ISAC beamforming problem as a semidefinite program (SDP), then derive its Lagrangian dual and analyze the resulting structure, and finally construct an iterative algorithm that optimizes the Lagrange multipliers via subgradient ascent.

\subsection{SDP Reformulation}

To address the nonconvexity of problem \eqref{eq:jsc_exp}, we apply SDR by replacing $\v v_k \v v_k^\hermitian$ with $\m V_k \succeq \m 0$ and dropping the rank-one constraint.
As shown in \cite[Lemma~1 and Appendix~A]{attiah2025uplink-downlink}, this relaxation is tight for problem \eqref{eq:jsc_exp}, yielding the equivalent formulation
\begin{equation}
	\begin{aligned}
		\min_{\{\m V_k \succeq 0\}} &\quad \sum_{k\in\cK}  \tr(\m V_k)\\
		\st~~ &\; \bigl(1+\tfrac{1}{\gamma_k}\bigr) \v h_k^\hermitian \m V_k \v h_k \geq \sum_{j\in\cK} \v h_k^\hermitian \m V_j \v h_k + \sigma_k^2, \fk,\\
		&\; \sum_{q\in\cQ} \bigl\langle \sum_{k\in\cK} \m V_k, \m M_q \bigr\rangle^2 \leq Q \eta.
	\end{aligned}
	\label{eq:sdr}
\end{equation}
Note that the sensing beamforming matrix $\m V_s$ no longer appears as a separate variable, since the contribution of $\m V_s \m V_s^\hermitian$ to the transmit covariance is absorbed into the relaxed matrices $\{\m V_k \succeq \m 0\}$.
Applying the Schur complement to the MSE constraint in problem \eqref{eq:sdr}, we obtain the following equivalent SDP reformulation of problem \eqref{eq:jsc_exp}:
\begin{equation}
	\begin{aligned}
		\min_{\{\m V_k \succeq 0\}} &\quad \sum_{k\in\cK} \tr(\m V_k)\\
		\st~~ &\; \bigl(1+\tfrac{1}{\gamma_k}\bigr) \v h_k^\hermitian \m V_k \v h_k \geq \sum_{j\in\cK} \v h_k^\hermitian \m V_j \v h_k + \sigma_k^2, \fk,\\
		&\; \begin{bmatrix}
			\m I & \cM (\sum_{k \in \cK} \m V_k) \\
			(\cM (\sum_{k \in \cK} \m V_k))^\transpose &  Q \eta
		\end{bmatrix} \succeq \m 0,
	\end{aligned}
	\label{eq:sdp}
\end{equation}
where $ \cM: \m V \rightarrow [\langle \m V, \m M_1 \rangle, \langle \m V, \m M_2 \rangle, \d, \langle \m V, \m M_Q \rangle]^\transpose$.

\subsection{Lagrangian Dual and Analysis}

Let the following matrix denote the dual variables associated with the sensing constraint in \eqref{eq:sdp}:
\[
\begin{bmatrix}
	\m \Lambda & \v \lambda \\
	\v \lambda^\transpose &  \mu
\end{bmatrix} \succeq \m 0
\]
with $\m \Lambda \in \mathbb R^{Q\times Q}, \v \lambda = [\lambda_1, \lambda_2, \d, \lambda_Q]^\transpose \in \mathbb R^{Q}$, and $\mu \in \mathbb R$.
Then, the dual of problem \eqref{eq:sdp} is given by
\begin{equation}
	\begin{aligned}
		\max_{\m \Lambda \succeq \mu^{-1} \v \lambda \v \lambda^\transpose, \v \lambda \in \mathbb R^{Q}, \mu \geq 0} \quad d(\v \lambda) - \tr(\m \Lambda)  - \mu Q \eta,
	\end{aligned}
	\label{eq:dual}
\end{equation}
where $d(\v \lambda)$ is the objective value of
\begin{equation}
	\begin{aligned}
		\min_{\{\m V_k \succeq 0\}} &\quad \tr\!\left(\m B(\v \lambda) \m R\right) \\
		\st~~&\; \bigl(1+\tfrac{1}{\gamma_k}\bigr) \v h_k^\hermitian \m V_k \v h_k \geq \sum_{j\in\cK} \v h_k^\hermitian \m V_j \v h_k + \sigma_k^2, \fk \\
	\end{aligned}
	\label{eq:gdb}
\end{equation}
with 
\begin{equation}
\m B(\v \lambda) = \m I - 2 \sum_{q \in \cQ} \lambda_q \m M_q
\label{eq:B_lambda}
\end{equation}

Problem \eqref{eq:gdb} is a classical DB problem except that the weighting matrix $\m B(\v \lambda)$ in \eqref{eq:B_lambda} could be semidefinite or even indefinite.
When $\m B(\v \lambda)$ is positive definite, existing fixed point iteration methods can be directly applied \cite{rashidDL1998}.
However, when $\m B(\v \lambda)$ is semidefinite or indefinite, existing algorithms and analyses are no longer applicable.
We refer to problem \eqref{eq:gdb} with such a weighting matrix as the GDB problem.
The GDB problem commonly arises in duality-based approaches \cite{attiah_beamforming_2024,attiah2025uplink-downlink,zhang_joint_2025}, although those works consider different problem settings and formulations.
A detailed theoretical study of the GDB problem will be presented in Section~\ref{sec:inner}.

Notice that the optimal solution of problem \eqref{eq:dual} must satisfy
\[
\m \Lambda = \mu^{-1} \v \lambda \v \lambda^\transpose, \quad \mu = (Q\eta)^{-1/2}\|\v \lambda\|.
\]
Hence, by plugging this into problem \eqref{eq:dual}, the dual problem is simplified to
\begin{equation}
	\begin{aligned}
		\max_{\v \lambda \in \mathbb R^{Q}} \quad \tilde d(\v \lambda) := d(\v \lambda) - 2 \sqrt{Q\eta} \| \v \lambda \|.
	\end{aligned}
	\label{eq:outer}
\end{equation}
When $\v \lambda \neq \v 0$, the objective function in \eqref{eq:outer} is differentiable, and its gradient is given by
\begin{equation}
\nabla \tilde d(\v \lambda) = - 2 \cM \left(\sum_{k \in \cK} \m V_k^\star(\v \lambda)\right) - \frac{2 \sqrt{Q\eta} \v \lambda}{\|\v \lambda\|},
\end{equation}
where $\{\m V_k^\star(\v \lambda)\}$ is an optimal solution to problem \eqref{eq:gdb}.
When $\v \lambda = \v 0$, the term $\|\v \lambda\|$ renders $\tilde d$ nondifferentiable, and the optimality condition reduces to $\|\nabla d(\v 0)\| \leq 2\sqrt{Q\eta}$, where $\nabla d(\v 0) = -2\cM\!\left(\sum_{k \in \cK} \m V_k^\star(\v 0)\right)$ is the gradient of the smooth part $d(\v \lambda)$ at $\v \lambda = \v 0$.
Accordingly, we define the optimality residual as $r(\v \lambda) := \|\nabla \tilde d(\v \lambda)\|$ for $\v \lambda \neq \v 0$ and $r(\v 0) := \max\!\left(0,\, \|\nabla d(\v 0)\| - 2\sqrt{Q\eta}\right)$, and use it as the stopping criterion in Algorithm~\ref{alg:dual_algorithm}.
Notice that when $\eta$ is sufficiently large, the optimal solution to problem \eqref{eq:outer} is $\v \lambda = \v 0$, i.e., the sensing constraint does not affect problem \eqref{eq:sdp}.

\begin{algorithm}[t]
\caption{Duality-Based Algorithm for Solving Problem \eqref{eq:outer}}
\label{alg:dual_algorithm}
\begin{algorithmic}[1]
\REQUIRE Initial $\v \lambda_0$, step sizes $\{\alpha_t\}$, backtracking factor $\tau \in (0, 1)$, tolerance $\epsilon$.
\FOR{$t = 0, 1, 2, \d$}
    \STATE Solve GDB problem \eqref{eq:gdb} with $\v \lambda_t$ to obtain $\{\m V_{k,t}\}$.
    \STATE Compute the optimality residual $r(\v \lambda_t)$.
    \IF{ $r(\v \lambda_t) \leq \epsilon$}
   	\STATE \textbf{break}.
    \ENDIF
    \STATE Find the smallest $\ell \geq 0$, denoted as $\ell_t$, such that GDB problem \eqref{eq:gdb} with $\v \lambda = \v \lambda_t + \alpha_t \tau^{\ell} \nabla \tilde d(\v \lambda_t)$ is bounded.
    \STATE Update $\v \lambda_{t+1} = \v \lambda_t + \alpha_t \tau^{\ell_t} \nabla \tilde d(\v \lambda_t)$.
\ENDFOR
\STATE \textbf{Output:} Beamforming matrices $\{\m V_{k,t}\}$.
\end{algorithmic}
\end{algorithm}
\vspace{-8pt}

\subsection{Duality-Based Algorithm for Solving Problem \eqref{eq:jsc_exp}}

We now propose an efficient duality-based algorithm for solving problem \eqref{eq:outer}, which is equivalent to solving problem \eqref{eq:sdp} by strong duality \cite{boyd2004ConvexOptimization}.
The key insight is that problem \eqref{eq:outer} can be viewed as a maximization problem over the Lagrange multipliers $\v \lambda$, where the GDB problem \eqref{eq:gdb} with each fixed $\v \lambda$ is solved to compute $d(\v \lambda)$ and its first-order information.
More specifically, the algorithm, summarized as Algorithm \ref{alg:dual_algorithm}, alternates between solving the inner GDB problem for fixed dual variables and updating the dual variables using the first-order information.

\section{An FPI Algorithm for Solving Problem \eqref{eq:gdb}}\label{sec:inner}

In this section, we address the solution of the GDB problem \eqref{eq:gdb} under a fixed $\v\lambda$, where the dependency on $\v\lambda$ is omitted for notational brevity.
Our analysis proceeds as follows.
First, we derive the Lagrangian dual problem and establish its key theoretical properties.
Second, we develop an FPI algorithm for solving problem \eqref{eq:gdb} and describe the procedure for recovering the primal solution.
Third, we provide rigorous convergence and stability analysis, including the characterization of multiple fixed points that may arise in the GDB problem.
Finally, we analyze the convergence rates of the dual FPI \eqref{eq:fp_iter}, and establish the overall convergence guarantee for the proposed FPI algorithm.

\subsection{Lagrangian Dual and Its Properties}

\subsubsection{Dual Problem Formulation}
We first derive the Lagrangian dual problem of \eqref{eq:gdb}. Let $\v \beta = [\beta_1, \beta_2, \dots, \beta_K]^\transpose$ denote the dual variables associated with the SINR constraints in \eqref{eq:gdb}. The dual problem can be expressed as
\begin{equation}
\label{eq:dual2}
\begin{aligned}
\max_{\{\beta_k \geq 0\}} &\quad \sum_{k=1}^K \beta_k \sigma_k^2\\
\st~~&\quad
\m C(\v \beta)
\succeq
\frac{\beta_k(\gamma_k + 1)}{\gamma_k} \v h_k \v h_k^\hermitian, \fk,
\end{aligned}
\end{equation}
where $ \m C(\v \beta) = \m B + \sum_{j \in \cK} \beta_j \v h_j \v h_j^\hermitian$.
To reveal the underlying problem structure, we apply the Schur complement and reformulate \eqref{eq:dual2} into an equivalent form:
\begin{equation}
\label{eq:feas}
\begin{aligned}
\max_{\v \beta \in \mathcal F_1} &\quad \sum_{k=1}^K \beta_k \sigma_k^2\\
\st~&\quad \v \beta \leq I(\v \beta),
\end{aligned}
\end{equation}
where $I(\v \beta) = [I_1(\v \beta), I_2(\v \beta), \dots, I_K(\v \beta)]^\transpose$ with
\begin{equation}
I_k(\v \beta)
=
\frac{\gamma_k}{\gamma_k + 1}
\frac{1}{\v h_k^\hermitian \m C(\v \beta)^{-1} \v h_k}, \fk,
\label{eq:interference}
\end{equation}
and
\begin{multline}
\label{eq:check}
\mathcal F_1 = \{\v \beta \mid \v \beta \geq \v 0, \; \m C(\v \beta) \succeq \m 0, \\
\text{and } \v h_1, \v h_2, \dots, \v h_K \text{ lie in the range of } \m C(\v \beta)\}.
\end{multline}
Denote the feasible region of problem \eqref{eq:feas} as \begin{equation}
\mathcal F = \mathcal F_1 \cap \{\v \beta \mid \v \beta \leq I(\v \beta)\}.
\label{eq:feas_region}
\end{equation}
This formulation reveals a fixed point relationship between $\v \beta$ and $I(\v \beta)$, which forms the foundation for both the boundedness analysis and the design of the dual FPI for solving the dual problem.

\subsubsection{Properties of the Dual Problem}
We now establish a theoretical characterization of the dual problem, focusing on the existence, uniqueness, and fixed point properties of its solution.

\begin{proposition}[Properties of the Dual Problem]\label{prop:dual_characterization}
The following statements hold:
\begin{enumerate}
    \item[(a)] If problem \eqref{eq:dual2} is feasible, then it admits a unique optimal solution $\v \beta^\star$, and $\v \beta^\star \geq \v \beta$ for all $\v \beta \in \mathcal F$.
    \item[(b)] The optimal solution $\v \beta^\star$ of problem \eqref{eq:dual2} is a fixed point of $I(\cdot)$ in \eqref{eq:interference}.
    \item[(c)] Problem \eqref{eq:gdb} is bounded if and only if there exists $\v \beta \in \mathcal F_1$ such that $\v \beta = I(\v \beta)$.
\end{enumerate}
\end{proposition}

\begin{proof}
The proof consists of three parts corresponding to (a), (b), and (c), respectively.

(a) To prove the existence and uniqueness, define $\bar{\v \beta} = [\bar \beta_1, \bar \beta_2, \d, \bar \beta_K]^\transpose$ with
$\bar \beta_k = \sup \{\beta_k \mid \v \beta \in \mathcal F\}$.
It follows that $\m C(\bar{\v \beta}) \succeq \m 0$ and $\v h_1, \v h_2, \dots,\v h_K$ lie in the range of $\m C(\bar{\v \beta})$.
By definition, $\v \beta \leq \bar{\v \beta}$ for any $\v \beta \in \mathcal F$.
The monotonicity of $I(\cdot)$ in \eqref{eq:interference} implies $\v \beta \leq I(\v \beta) \leq I(\bar{\v \beta})$.
Taking the supremum on both sides gives $\bar{\v \beta} \leq I(\bar{\v \beta})$, i.e., $\bar{\v \beta} \in \mathcal F$.
Thus, $\bar{\v \beta}$ is the desired unique maximal element $\v \beta^\star$.

(b) To prove the fixed point condition, suppose for contradiction that $\beta^\star_k < I_k(\v \beta^\star)$ for some $k$.
Let $\tilde \beta_k = \beta^\star_k + \epsilon$ and $\tilde \beta_j = \beta^\star_j$ for $j \neq k$.
By the monotonicity and continuity of $I_k(\cdot)$, there exists $\epsilon > 0$ such that $\tilde \beta_k < I_k(\tilde{\v \beta})$.
The new $\tilde{\v \beta}$ yields a strictly larger objective while remaining feasible, contradicting the optimality.
Hence, $\v \beta^\star = I(\v \beta^\star)$.

(c) If the primal problem is bounded, the dual problem is feasible by strong duality, and its optimal solution satisfies the fixed point condition.
Conversely, if a fixed point $\v \beta \in \mathcal F_1$ exists, it is feasible for \eqref{eq:dual2}, implying the boundedness of problem \eqref{eq:gdb} by weak duality.
\end{proof}

Proposition \ref{prop:dual_characterization} (a) and (b) provide an important characterization of the dual solution, which is instrumental in the subsequent algorithmic development and theoretical analysis.
Proposition \ref{prop:dual_characterization} (c) establishes a necessary and sufficient condition for the boundedness of problem \eqref{eq:gdb}, which serves as a feasibility criterion for the current iterate in Line 7 of Algorithm \ref{alg:dual_algorithm}.
The work \cite{attiah2025uplink-downlink} also studied the boundedness of problem \eqref{eq:gdb}, referred to as ``admissibility'' therein, but does not specify how to verify it.
Proposition \ref{prop:dual_characterization} (c) provides computable conditions for this purpose.

\subsection{FPI Algorithm for Solving Problem \eqref{eq:gdb}}\label{subsec:FPI}

In this subsection, we first describe the FPI algorithm for solving the dual problem and the procedure for recovering the primal solution assuming convergence to the optimal dual variable $\v \beta^\star$.
The rigorous convergence guarantee will be established in Section \ref{subsec:convergence_stability}.

\subsubsection{FPI of the Dual Problem}
Proposition~\ref{prop:dual_characterization}~(b) motivates the following FPI:
\begin{equation}
\label{eq:fp_iter}
\v \beta^{(i+1)} = I(\v \beta^{(i)}),
\end{equation}
starting from an initial point $\v \beta^{(0)}$ to solve problem \eqref{eq:dual2}.
While prior work has established the convergence of \eqref{eq:fp_iter} to the unique fixed point when $\m B \succ \m 0$, the general case where $\m B$ is semidefinite or indefinite presents significant challenges. In such scenarios, multiple fixed points may exist (see Fig.~\ref{fig:alg_behaviour} and Appendix~\ref{apd:multiple_fp_example} for an illustrative example), and the convergence to the correct optimal $\v \beta^\star$ depends critically on the initial point $\v \beta^{(0)}$.

\subsubsection{Recovery of the Primal Solution}
Upon obtaining the optimal dual solution $\v \beta^\star$, we recover the primal solution $\{\m V_k^\star\}$ using the KKT conditions.
By Slater's condition, there exists a strictly feasible point $\hat{\v \beta} \in \mathcal F$ such that $\m C(\hat{\v \beta}) \succ \m 0$.
Since $\v \beta^\star$ dominates all feasible points by Proposition~\ref{prop:dual_characterization}~(a), we have $\v \beta^\star \geq \hat{\v \beta}$, which implies $\m C(\v \beta^\star) \succeq \m C(\hat{\v \beta}) \succ \m 0$.
Combining this with the complementary slackness condition yields
\begin{equation*}
\tr \!\left(
\m V_k^\star \Big(\m C(\v \beta^\star)
- \tfrac{\beta_k^\star (\gamma_k + 1)}{\gamma_k} \v h_k \v h_k^\hermitian \Big)
\right) = 0, \fk.
\end{equation*}
This condition implies that each $\m V_k$ must be rank-one.
Let $\tilde{\v v}_k = {\m C(\v \beta^\star)^{-1} \v h_k}/{\|\m C(\v \beta^\star)^{-1} \v h_k\|}$ and $\m G_{k,j} = |\v h_k^\hermitian \tilde{\v v}_j|^2$. Then $\m V_k^\star = p_k^\star \tilde{\v v}_k \tilde{\v v}_k^\hermitian$ with $p_k^\star > 0$ satisfying
\begin{equation}
\left(1+\frac{1}{\gamma_k}\right) p_k^\star \m G_{k, k} = \sum_{j\in\cK} p_j^\star \m G_{k, j} + \sigma_k^2, \fk,
\label{eq:pk_system}
\end{equation}
which constitutes a system of linear equations in $\{p_k^\star\}$. 
Equation \eqref{eq:pk_system} motivates the following primal FPI:
\begin{equation}
p_k^{(i+1)} = \frac{\gamma_k}{\gamma_k + 1} \frac{1}{\m G_{k, k}} \left( \sum_{j \neq k} p_j^{(i)} \m G_{k, j} + \sigma_k^2 \right), \fk.
\label{eq:primal_fpi}
\end{equation}

We write \eqref{eq:pk_system} in matrix form as $\m P \v p = \v b$, where $\m P \in \mathbb{R}^{K \times K}$ has diagonal entries $P_{k,k} = (1+1/\gamma_k) \m G_{k,k}$ and off-diagonal entries $P_{k,j} = -\m G_{k,j}$ for $j \neq k$, and $\v b = [\sigma_1^2, \sigma_2^2, \ldots, \sigma_K^2]^\transpose$ is the noise power vector.
The primal FPI \eqref{eq:primal_fpi} can then be written as $\v p^{(i+1)} = \m D^{-1}\m P_{\mathrm{off}} \v p^{(i)} + \m D^{-1}\v b$, where $\m D$ and $\m P_{\mathrm{off}}$ contain the diagonal and off-diagonal entries of $\m P$, respectively.
We have the following convergence result.

\begin{lemma}[Linear Convergence of the Primal FPI]\label{lem:primal_convergence_rate}
If \eqref{eq:pk_system} admits a positive solution $\v p^\star > \v 0$, then the primal FPI \eqref{eq:primal_fpi} converges to $\v p^\star$ from any nonnegative initialization.
Furthermore, the convergence is linear with the rate $\rho(\m D^{-1}\m P_{\mathrm{off}})$, where $\rho(\cdot)$ denotes the spectral radius of the iteration matrix.
\end{lemma}

\begin{proof}
The iteration $\v p^{(i+1)} = \m D^{-1}\m P_{\mathrm{off}} \v p^{(i)} + \m D^{-1}\v b$ is an affine mapping with nonnegative iteration matrix $\m D^{-1}\m P_{\mathrm{off}} \geq \m 0$ and nonnegative constant vector $\m D^{-1}\v b \geq \v 0$.
Hence, the iteration defines a standard interference function \cite{yates1995FrameworkUplinkPower,foschini1993SimpleDistributed}.
By the standard interference function theory, if a positive fixed point $\v p^\star$ exists, then the iteration converges to $\v p^\star$ from any nonnegative initialization.
Since $\v p^\star$ exists by assumption, convergence follows, and hence $\rho(\m D^{-1}\m P_{\mathrm{off}}) < 1$, which characterizes the linear convergence rate for fixed point iterations. 
\end{proof}

\subsubsection{Proposed FPI Algorithm}

\begin{algorithm}[t]
\caption{Proposed FPI Algorithm for Solving Problem~\eqref{eq:gdb}}
\label{alg:fp_algorithm}
\begin{algorithmic}[1]
\REQUIRE Initial point $\v \beta^{(0)}$.
\STATE Obtain $\v \beta^\star$ by performing the iteration in \eqref{eq:fp_iter} until convergence.
\IF{The iteration converges and $\v \beta^\star \in \mathcal F_1$ in \eqref{eq:check}}
    \STATE Obtain $\{p_k^\star\}$ by performing the iteration in \eqref{eq:primal_fpi} until convergence.
    \STATE Recover beamforming matrices $\m V_k^\star = p_k^\star \tilde{\v v}_k \tilde{\v v}_k^\hermitian$.
    \STATE \textbf{Return:} $\{\m V_k^\star\}$.
\ELSE
    \STATE \textbf{Return:} Unbounded problem.
\ENDIF
\end{algorithmic}
\end{algorithm}
\vspace{-8pt}

The proposed FPI algorithm for solving \eqref{eq:gdb} is summarized in Algorithm~\ref{alg:fp_algorithm}.
The algorithm consists of two components: the dual FPI \eqref{eq:fp_iter} for solving the dual problem, and the primal FPI \eqref{eq:primal_fpi} for recovering the primal solution.
While the primal FPI has straightforward convergence guarantees due to its linear nature, the key theoretical challenge lies in analyzing the dual FPI \eqref{eq:fp_iter}, which exhibits a more complicated behavior when the weighting matrix $\m B$ in \eqref{eq:B_lambda} is semidefinite or indefinite.
In the following subsection, we establish the theoretical foundations of the dual FPI, including its convergence properties and stability characterization.

\subsection{Convergence and Stability Analysis of the Dual FPI}\label{subsec:convergence_stability}

In this subsection, we analyze the convergence and stability properties of the dual FPI \eqref{eq:fp_iter}.
Unlike the classical DB problem where the weighting matrix is positive definite, the GDB problem with semidefinite or indefinite $\m B$ in \eqref{eq:B_lambda} admits a nontrivial feasible region and may have multiple fixed points of the interference function $I(\cdot)$; see Appendix~\ref{apd:multiple_fp_example} for an illustrative example.
As a result, the correspondence between fixed points and the optimal solution is no longer one-to-one, and the stability properties of different fixed points become crucial for understanding the behavior of the dual FPI.

\subsubsection{Convergence Guarantee}

The following theorem establishes the convergence of the dual FPI \eqref{eq:fp_iter}.

\begin{theorem}\label{thm:convergence}
Suppose that problem \eqref{eq:gdb} is bounded and Slater's condition holds for problem \eqref{eq:dual2}, i.e., there exists $\hat{\v \beta} \in \mathcal F$ such that $\m C(\hat{\v \beta}) \succ \hat{\beta}_k (\gamma_k + 1)\gamma_k^{-1} \v h_k \v h_k^\hermitian$ holds for all $k \in \cK$.
Let $\hat{\mathcal S} = \{\v \beta \mid \v \beta \geq \hat{\v \beta}\}$.
Then, the dual FPI \eqref{eq:fp_iter} converges to the optimal dual solution $\v \beta^\star$ for any initialization $\v \beta^{(0)} \in \hat{\mathcal S}$.
Moreover, $\v \beta^\star$ is the unique fixed point of $I(\cdot)$ within $\hat{\mathcal S}$.
\end{theorem}

\begin{proof}
Define the shifted variable $\v \delta = \v \beta - \hat{\v \beta}$, and consider the function $J(\v \delta) = I(\v \beta) - \hat{\v \beta} = I(\hat{\v \beta} + \v \delta) - \hat{\v \beta}$.
We verify that $J(\cdot)$ is a standard interference function \cite{yates1995FrameworkUplinkPower}:
\begin{enumerate}
    \item $J(\v 0) = I(\hat{\v \beta}) - \hat{\v \beta} > \v 0$;
    \item $J(\cdot)$ is monotone due to the monotonicity of $I(\cdot)$;
    \item for any $\alpha > 1$, the concavity of $I(\cdot)$ \cite{boche2008ConcaveConvex} on $\mathcal F$ implies
    \[
    J(\alpha \v \delta) < \alpha J(\v \delta),
    \]
    ensuring scalability. 
\end{enumerate}
The fixed point of $J(\cdot)$ is $\v \delta^\star = \v \beta^\star - \hat{\v \beta}$, which satisfies $J(\v \delta^\star) = \v \delta^\star$.
By the standard interference function theory, the iteration $\v \delta^{(i+1)} = J(\v \delta^{(i)})$ converges to $\v \delta^\star$ from any initialization $\v \delta^{(0)} \in \mathbb{R}_+^K$.
Mapping back to the original variable, this implies $\v \beta^{(i+1)} = I(\v \beta^{(i)})$ converges to $\v \beta^\star = \hat{\v \beta} + \v \delta^\star$ for any initialization $\v \beta^{(0)} \in \hat{\mathcal S}$.
\end{proof}

Some remarks on Theorem~\ref{thm:convergence} are in order.
First, although the analysis is formulated in terms of $\hat{\v \beta}$, the proposed FPI algorithm itself does not depend on this auxiliary variable, i.e., $\hat{\v \beta}$ serves purely for theoretical purposes to characterize the convergent initialization region.
Second, the convergent initialization region can equivalently be characterized without explicitly referencing $\hat{\v \beta}$:
\begin{equation}
\label{eq:S}
\mathcal S = \bigcup \left\{ \hat{\mathcal S} \mid \hat{\v \beta} \in \mathcal F,~ \hat{\v \beta} < I(\hat{\v \beta}) \right\}.
\end{equation}
In particular, when $\m B \succ \m 0$, one can choose $\hat{\v \beta} = \v 0$, yielding $\hat{\mathcal S} = \mathbb{R}_+^K$ and recovering the classical convergence result as a special case. Theorem~\ref{thm:convergence} generalizes this convergent initialization region to arbitrary $\m B$.

\subsubsection{Stability Analysis}

As mentioned earlier, the potential existence of multiple fixed points introduces ambiguity in determining which fixed point corresponds to the optimal solution.
Recall from Proposition~\ref{prop:dual_characterization} that the maximal fixed point, denoted as $\v \beta^\star$, exists and coincides with the optimal dual solution.
We now investigate the stability properties of the fixed points of the mapping $I(\cdot)$ under the dual FPI \eqref{eq:fp_iter}.
In particular, we show that the maximal fixed point is the unique stable fixed point, whereas all other fixed points—if they exist—are unstable.

\begin{theorem}[Stability Characterization of Fixed Points]
\label{thm:stability_characterization}
Assume that the conditions of Theorem~\ref{thm:convergence} hold.  
Then the maximal fixed point $\v \beta^\star$ is stable, whereas any other fixed point of $I(\cdot)$ is unstable.
\end{theorem}

\begin{proof}
We first establish the stability of $\v \beta^\star$.
By Theorem~\ref{thm:convergence}, there exists $\hat{\v \beta} \in \mathcal F$ such that any initialization satisfying $\v \beta^{(0)} \ge \hat{\v \beta}$ leads to the convergence to $\v \beta^\star$.
Since $\hat{\v \beta} < I(\hat{\v \beta}) \le \v \beta^\star$ holds component-wise, $\v \beta^\star$ lies in the interior of $\hat{\mathcal S} = \{\v \beta \mid \v \beta \geq \hat{\v \beta}\}$.
Therefore, there exists an open ball around $\v \beta^\star$ entirely contained in $\hat{\mathcal S}$, and by Theorem~\ref{thm:convergence} all trajectories starting from this ball converge to $\v \beta^\star$, establishing the stability.

We next prove the instability of any non-maximal fixed point.
Let $\v \beta \neq \v \beta^\star$ be another fixed point of $I(\cdot)$.
By the maximality of $\v \beta^\star$, we have $\v \beta \le \v \beta^\star$.
Recall that $\hat{\v \beta} \in \mathcal F$ is the point guaranteed by Theorem~\ref{thm:convergence}.
Consider the convex combination
\[
\v \beta(t) = (1-t)\v \beta + t \hat{\v \beta},
\quad t \in (0,1].
\]
Since $\m C(\v \beta(t)) = (1-t)\m C(\v \beta) + t \m C(\hat{\v \beta})$ and $\m C(\hat{\v \beta}) \succ \m 0$, we have $\v \beta(t) \in \mathcal F_1$ for all $t \in (0,1]$.
Moreover, by the concavity of $I(\cdot)$,
\[
I(\v \beta(t)) \geq (1-t) I(\v \beta) + t I(\hat{\v \beta}) > (1-t) \v \beta + t \hat{\v \beta} = \v \beta(t),
\]
where the strict inequality follows from $\hat{\v \beta} < I(\hat{\v \beta})$ and $t > 0$.
Hence $\v \beta(t)$ satisfies the strict feasibility condition in Theorem~\ref{thm:convergence}, and the iteration from $\v \beta(t)$ converges to $\v \beta^\star$.
Since $\v \beta(t)$ can approach $\v \beta$ arbitrarily closely as $t \to 0$, trajectories starting near $\v \beta$ escape to $\v \beta^\star$.
Therefore, $\v \beta$ cannot attract nearby trajectories and is unstable.
\end{proof}

Theorem~\ref{thm:stability_characterization} characterizes the stability properties of fixed points in the GDB problem.
Even when multiple fixed points exist, the theorem shows that only the maximal fixed point is stable, whereas all other fixed points are unstable.
This theoretical insight explains the numerical behavior observed in Section~\ref{sec:numerical}.
Specifically, when the FPI is initialized within the convergent initialization region $\mathcal S$ from Theorem~\ref{thm:convergence}, it consistently converges to $\v \beta^\star$, ensuring the algorithm's robustness and reliability in practice.

In summary, Theorem~\ref{thm:convergence} generalizes classical convergence guarantees to semidefinite and indefinite weighting matrices by explicitly characterizing the admissible initialization region $\mathcal S$, and Theorem~\ref{thm:stability_characterization} further shows that the maximal fixed point is uniquely stable. Together, these results address both convergence and stability in the presence of multiple fixed points.
We next analyze the convergence rates of the algorithm in Section~\ref{subsec:convergence_rate}.

\subsection{Convergence Rate Analysis of the Dual FPI}\label{subsec:convergence_rate}

In this subsection, we analyze the convergence rate of the dual FPI \eqref{eq:fp_iter}.
The convergence rate of the primal FPI \eqref{eq:primal_fpi} has been established in Lemma~\ref{lem:primal_convergence_rate}.
Here we establish the linear convergence rate for the dual FPI and discuss how the convergence behavior depends on problem parameters.

We first analyze the convergence rate of the dual FPI \eqref{eq:fp_iter} for solving the dual problem \eqref{eq:dual2}.
Recall from the proof of Theorem~\ref{thm:convergence} that the dual FPI can be expressed in terms of the shifted variable $\v{\delta}^{(i)} = \v{\beta}^{(i)} - \hat{\v{\beta}}$ as $\v{\delta}^{(i+1)} = J(\v{\delta}^{(i)})$, where $J(\v{\delta}) = I(\hat{\v{\beta}} + \v{\delta}) - \hat{\v{\beta}}$ with fixed point $\v{\delta}^\star = \v{\beta}^\star - \hat{\v{\beta}}$.
The convergence analysis in this subsection is formulated in terms of $\v{\delta}^{(i)}$ for notational convenience.

To characterize the convergence rate, we follow the approach in \cite{nuzman2007ContractionApproachPower} and define the metric $\mu: \mathbb{R}_{++}^K \times \mathbb{R}_{++}^K \rightarrow \mathbb{R}_+$ as
\begin{equation}
\mu(\v y, \v z) = \max_{k \in \cK} \left| \log \left( \frac{y_k}{z_k} \right) \right|,
\label{eq:metric}
\end{equation}
where $\mathbb{R}_{++}^K$ denotes the set of strictly positive $K$-dimensional vectors.

\begin{theorem}[Linear Convergence Rate of the Dual FPI]
\label{thm:fpi_convergence_rate}
Under the conditions of Theorem~\ref{thm:convergence}, the dual FPI \eqref{eq:fp_iter} converges linearly to the optimal dual solution $\v \beta^\star$.
Let $\hat{\m B} = \m B + \sum_{j \in\cK} \hat{\beta}_j \v h_j \v h_j^\hermitian = \m C(\hat{\v \beta})$ denote the positive definite matrix determined by the strictly feasible point $\hat{\v \beta}$ from Slater's condition.
The asymptotic convergence rate under the metric $\mu(\cdot, \cdot)$ satisfies
\begin{equation}
\limsup_{i\rightarrow \infty}\frac{\mu(\v{\delta}^{(i+1)}, \v{\delta}^\star)}{\mu(\v{\delta}^{(i)}, \v{\delta}^\star)} \leq \frac{\lambda}{1 + \lambda},
\label{eq:c_r_limit}
\end{equation}
where 
\begin{equation}
\lambda = \rho(\hat{\m B} \sum_{j\in\cK} \delta_j^\star \v h_j \v h_j^\hermitian).
\label{eq:lambda_def}
\end{equation}
\end{theorem}

\begin{proof}
See Appendix~\ref{apd:fpi_convergence_rate}.
\end{proof}

The asymptotic convergence rate in Theorem~\ref{thm:fpi_convergence_rate} is characterized by ${\lambda}/{(1 + \lambda)}$, where $\lambda$ in \eqref{eq:lambda_def} combines the PSD matrix $\hat{\m B}$ determined by the strictly feasible point $\hat{\v \beta}$ from Slater's condition with the component $\sum_{j\in\cK} \delta_j^\star \v h_j \v h_j^\hermitian$ determined by the optimal dual solution $\v \beta^\star = \hat{\v \beta} + \v \delta^\star$.
In particular, when the SINR targets $\{\gamma_k\}$ are high and the problem approaches the boundary of feasibility, $\v \delta^\star$ increases, leading to a larger $\lambda$ and hence slower convergence.
These theoretical insights will be verified through numerical experiments in Section~\ref{sec:numerical}.

\subsection{Overall Algorithm Convergence}

Having established the convergence, stability, and convergence rate properties of the inner FPI in Algorithm~\ref{alg:fp_algorithm}, we now analyze the overall convergence of Algorithm~\ref{alg:dual_algorithm}.
Recall that Algorithm~\ref{alg:dual_algorithm} employs a backtracking line search to ensure the boundedness of the GDB problem at each outer iterate.
The boundedness is verified via Proposition~\ref{prop:dual_characterization}~(c) by checking whether Algorithm~\ref{alg:fp_algorithm} converges to a fixed point $\v \beta^\star$ satisfying $\m C(\v \beta^\star) \succ \m 0$.

\begin{theorem}[Convergence of the Dual-FPI Algorithm]
\label{thm:overall_convergence}
Consider Algorithm~\ref{alg:dual_algorithm} for solving problem~\eqref{eq:outer}, where the inner GDB problem is solved by Algorithm~\ref{alg:fp_algorithm} and the boundedness check in the backtracking is performed via Proposition~\ref{prop:dual_characterization}~(c).
Suppose the following conditions hold:
\begin{enumerate}
	\item[(a)] Problem~\eqref{eq:outer} admits an optimal solution $\v \lambda^\star$, and the subgradients are uniformly bounded: $\|\nabla \tilde d(\v \lambda_t)\| \leq G$ for some $G > 0$.
	\item[(b)] At each outer iteration, Algorithm~\ref{alg:fp_algorithm} is initialized with $\v \beta^{(0)} \in \mathcal S$, where $\mathcal S$ is defined in \eqref{eq:S}.
	\item[(c)] The effective step sizes $\bar \alpha_t := \alpha_t \tau^{\ell_t}$ satisfy $\sum_{t=0}^{\infty} \bar \alpha_t = \infty$ and $\sum_{t=0}^{\infty} \bar \alpha_t^2 < \infty$.
\end{enumerate}
Then Algorithm~\ref{alg:dual_algorithm} is well-defined, and the sequence $\{\v \lambda_t\}$ satisfies
\begin{equation}
\label{eq:overall_rate}
\tilde d^\star - \max_{0 \leq s \leq t} \tilde d(\v \lambda_s) \leq \frac{\|\v \lambda_0 - \v \lambda^\star\|^2 + G^2 \sum_{s=0}^{t} \bar \alpha_s^2}{2 \sum_{s=0}^{t} \bar \alpha_s}.
\end{equation}
Furthermore, $\v \lambda_t \to \v \lambda^\star$, where $\v \lambda^\star$ is the unique maximizer of $\tilde d$.
\end{theorem}

\begin{proof}
By assumption~(b) and Theorem~\ref{thm:convergence}, Algorithm~\ref{alg:fp_algorithm} converges to the optimal dual solution $\v \beta^\star$ whenever the GDB problem is bounded.
Combining this with Proposition~\ref{prop:dual_characterization}~(c) ensures that the boundedness check in the backtracking is exact, i.e., a trial point is accepted if and only if the corresponding GDB problem is bounded.
Hence Algorithm~\ref{alg:dual_algorithm} is well-defined, and the inner solution at each accepted iterate is optimal.
Since $d(\v \lambda)$ is concave as the pointwise infimum of affine functions in $\v \lambda$ and $-2\sqrt{Q\eta}\|\v \lambda\|$ is concave, $\tilde d(\v \lambda)$ is concave, so that minimizing $-\tilde d(\v \lambda)$ is a convex optimization problem.
The convergence rate bound \eqref{eq:overall_rate} then follows from the standard subgradient method analysis for convex optimization \cite[Section~3.2]{boyd2003subgradient} under assumptions~(a) and~(c).
The iterate convergence under the uniqueness assumption of $\v \lambda^\star$ follows from the quasi-Fej\'{e}r monotonicity of $\{\|\v \lambda_t - \v \lambda^\star\|\}$.
\end{proof}

Assumption~(b) is addressed in practice by the initialization strategies in Section~\ref{sec:initialization}. In particular, the warm-start Method~II is guaranteed to lie in $\mathcal S$ when the outer step size is sufficiently small by Proposition~\ref{prop:warm_start}.
Assumption~(c) holds when the number of backtracking steps $\ell_t$ is uniformly bounded.
The convergence rate $O(1/\sqrt{t})$ implied by \eqref{eq:overall_rate} is the optimal rate for first-order methods applied to nonsmooth convex optimization \cite{nesterov2004introductory}.

\section{Initialization Strategies}\label{sec:initialization}

In this section, we discuss the initialization of the dual FPI \eqref{eq:fp_iter} within the Dual-FPI algorithm (i.e., Algorithm~\ref{alg:dual_algorithm}).
By Theorem~\ref{thm:convergence}, the~convergence to the optimal dual solution requires the initialization $\v \beta^{(0)}$ to lie in the convergent initialization region $\mathcal S$ defined in \eqref{eq:S}.
Constructing such an initialization requires finding a strict interior point $\hat{\v \beta} \in \mathcal F$ satisfying $\hat{\v \beta} < I(\hat{\v \beta})$.
While verifying the membership in $\mathcal S$ is computationally difficult since it requires the knowledge of the feasible region $\mathcal F$ in \eqref{eq:feas_region} of problem \eqref{eq:feas}, verifying the strict interior condition $\v \beta < I(\v \beta)$ is straightforward as it only involves evaluating the interference function.
Below, we present three initialization strategies that exploit this observation, each offering different trade-offs between theoretical guarantees and computational costs.

\subsection{Method I: SDP-Based Initialization}\label{subsec:sdp_init}

The most direct approach is to compute a strictly feasible point of the dual problem \eqref{eq:dual2} by solving the following semidefinite feasibility problem:
\begin{equation}\label{eq:sdp_slater}
\text{find}~~ \v \beta \geq \v 0 ~~\st~~ \m C(\v \beta) \succeq (1+\epsilon) \frac{\beta_k (\gamma_k + 1)}{\gamma_k} \v h_k \v h_k^\hermitian, \fk,
\end{equation}
where $\epsilon > 0$ is a small parameter enforcing strict feasibility.
Any solution $\hat{\v \beta}$ to \eqref{eq:sdp_slater} satisfies $\hat{\v \beta} < I(\hat{\v \beta})$, and thus any initialization $\v \beta^{(0)} \geq \hat{\v \beta}$ lies in $\mathcal S$.

Whenever problem \eqref{eq:gdb} is bounded, problem \eqref{eq:sdp_slater} is feasible and yields a valid initialization.
However, solving \eqref{eq:sdp_slater} requires solving an SDP, which incurs nontrivial computational overhead.

\subsection{Method II: Warm-Start Initialization}\label{subsec:warm_start}

The second method leverages the outer loop structure of Algorithm~\ref{alg:dual_algorithm} by using the optimal dual solution from the previous outer iteration as the initialization for the current iteration.
Let $\v \beta^\star_t$ denote the optimal dual solution obtained at the $t$-th outer iteration.
The warm-start strategy sets $\v \beta^{(0)} = \v \beta^\star_{t-1}$ for each outer iteration $t \geq 1$, avoiding the need to solve an SDP while exploiting the continuity of the optimal solution with respect to $\v \lambda$.
For the first outer iteration (i.e., $t = 0$), Method~I or Method~III can be used to obtain the initial $\v \beta^{(0)}$.

The following proposition establishes the theoretical validity of this warm-start strategy.

\begin{proposition}[Warm-Start Validity]\label{prop:warm_start}
When the outer step size is sufficiently small, the warm-start initialization $\v \beta^{(0)} = \v \beta^\star_{t-1}$ satisfies $\v \beta^{(0)} \in \mathcal S_t$, where $\mathcal S_t$ denotes the convergent initialization region for the $t$-th outer iteration.
\end{proposition}

\begin{proof}
Since $\v \beta^\star_{t-1} = I_{t-1}(\v \beta^\star_{t-1})$ and $I(\cdot)$ is continuous in $\v \lambda$, we have $I_t(\v \beta^\star_{t-1}) - \v \beta^\star_{t-1} = I_t(\v \beta^\star_{t-1}) - I_{t-1}(\v \beta^\star_{t-1}) \to \v 0$ as $\|\v \lambda_t - \v \lambda_{t-1}\| \to 0$.
For a sufficiently small step size, the continuity ensures $\v \beta^\star_{t-1} < I_t(\v \beta^\star_{t-1})$, so $\v \beta^\star_{t-1} \in \mathcal S_t$ by Theorem~\ref{thm:convergence}.
\end{proof}
Moreover, $\v \beta^\star_{t-1}$ is close to the new optimal solution $\v \beta^\star_t$, which significantly reduces the number of inner FPI iterations required for convergence.

\subsection{Method III: Heuristic Initialization}\label{subsec:heuristic_init}

For practical scenarios where computational simplicity is prioritized, we propose a closed-form heuristic initialization based on a single-user analysis.
Recall from \eqref{eq:interference} that the interference function $I_k(\v \beta)$ involves the matrix $\m C(\v \beta) = \m B + \sum_{j \in \cK} \beta_j \v h_j \v h_j^\hermitian$.
When the multi-user interference term $\sum_{j \in \cK} \beta_j \v h_j \v h_j^\hermitian$ is neglected, $\m C(\v \beta)$ reduces to $\m B$, and the fixed point equation $\beta_k = I_k(\v \beta)$ simplifies to $\beta_k = \gamma_k / ((\gamma_k + 1) \v h_k^\hermitian \m B^{-1} \v h_k)$.
This motivates the initialization
\begin{equation}\label{eq:heuristic_init}
\beta_k^{(0)} = c \cdot \frac{\gamma_k}{(\gamma_k + 1) \v h_k^\hermitian (\m B + \delta \m I)^{-1} \v h_k}, \fk,
\end{equation}
where $\delta = \max(0, -\lambda_{\min}(\m B)) + \epsilon$ with $\epsilon > 0$ is a regularization parameter that ensures $\m B + \delta \m I \succ \m 0$ when $\m B$ is indefinite.
The scalar $c > 1$ is a safety factor that compensates for the ignored multi-user interference.

The initialization \eqref{eq:heuristic_init} adapts to the problem structure by assigning larger values to users with higher SINR targets or weaker effective channel gains, consistent with the structure of the optimal solution.
When $\m B \succ \m 0$, the regularization $\delta$ reduces to a small perturbation $\epsilon$, and \eqref{eq:heuristic_init} approximates the single-user optimal dual variable.
The computational cost is dominated by a single matrix inversion, which is $O(M^3)$, making this method significantly cheaper than Methods I and II.
While this heuristic lacks a theoretical guarantee that $\v \beta^{(0)} \in \mathcal S$, it performs reliably in our numerical experiments across a wide range of problem configurations.

\section{Numerical Results}\label{sec:numerical}

In this section, we present numerical results to validate the proposed algorithms. We first examine the convergence behavior and initialization strategies, and then compare computational performance and solution quality with state-of-the-art approaches.

\subsection{Simulation Setup}

The channel vectors $\left\{\v h_k\right\}$ are generated using Rayleigh fading with unit variance.
For the sensing functionality, we consider radar beampattern matching with $Q = 36$ uniformly sampled angles $\left\{\theta_q\right\}$ in a $120^\circ$ sector.
The desired beampattern values $\left\{d(\theta_q)\right\}$ are drawn independently from a uniform distribution on $[0.5, 1.5]$.
The sensing performance is characterized by the MSE threshold $\eta \in [10^{-8}, 10^{-3}]$.
Communication requirements are specified through SINR targets $\gamma_k = \gamma$ dB for all users, with $\gamma$ uniformly drawn from $[-30, -10]$ dB. The noise power is normalized to $\sigma_k^2 = 1$ for all users.
These parameter ranges are selected to ensure problem feasibility.
For Algorithm~\ref{alg:dual_algorithm}, we set the initial dual variables $\v \lambda_0 = \v 0$, use alternate Barzilai--Borwein step sizes \cite{dai2005ProjectedBarzilaiBorweinMethods} with $\alpha_0 = 0.1$, backtracking factor $\tau = 0.5$, and convergence tolerance $\epsilon = 10^{-4}$.
For the inner FPI in Algorithm~\ref{alg:fp_algorithm}, we initialize $(\v \beta^{(0)})_k = 100$ for all $k \in \cK$ and terminate when $\|\v \beta^{(i+1)} - \v \beta^{(i)}\| < 10^{-12}$.

\subsection{Algorithm Behavior and Initialization Comparison}

\begin{figure}[t]
\centering
\begin{tabular}{cc}
\includegraphics[width=0.18\textwidth, trim=220 0 300 0, clip]{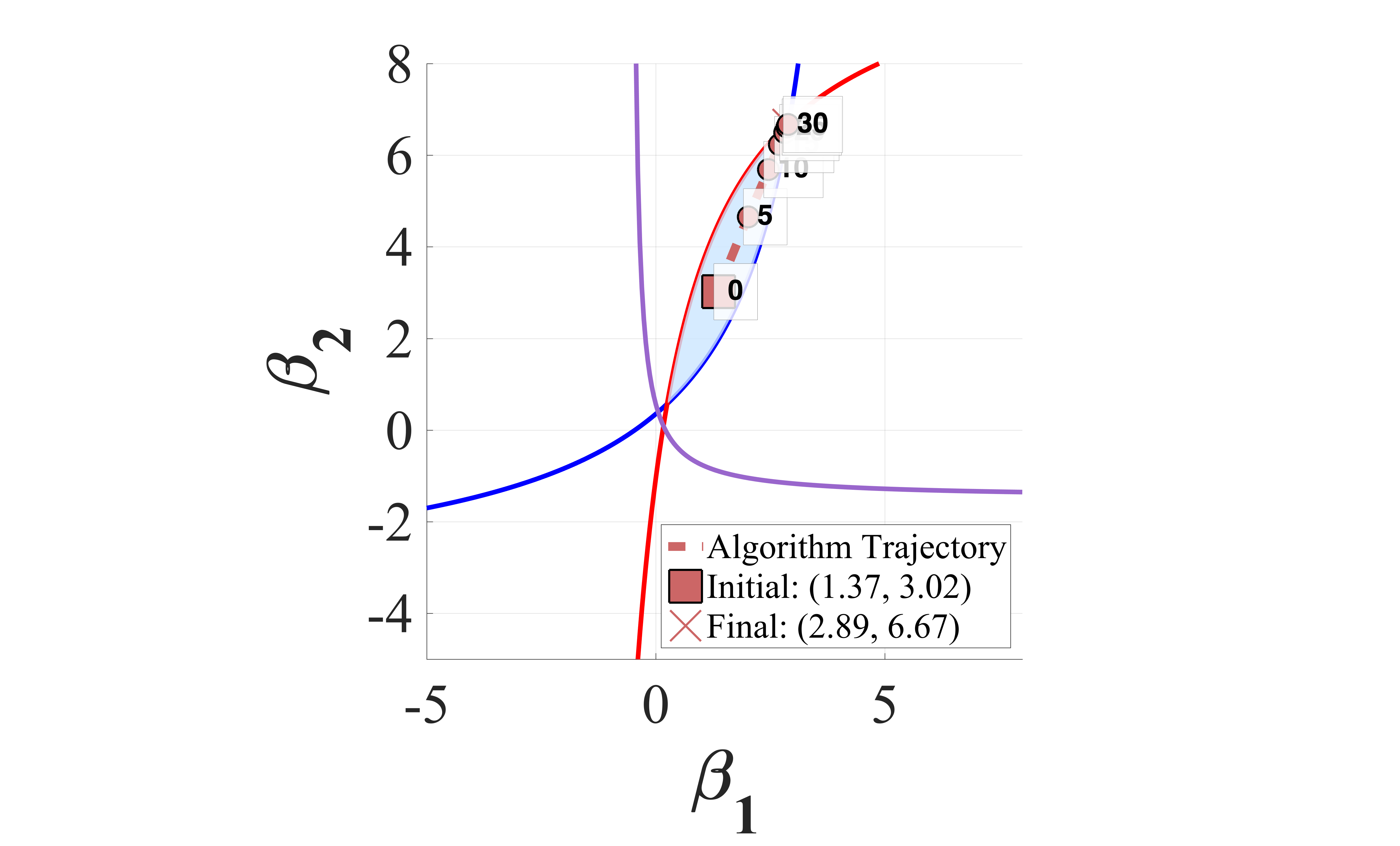} &
\includegraphics[width=0.18\textwidth, trim=220 0 300 0, clip]{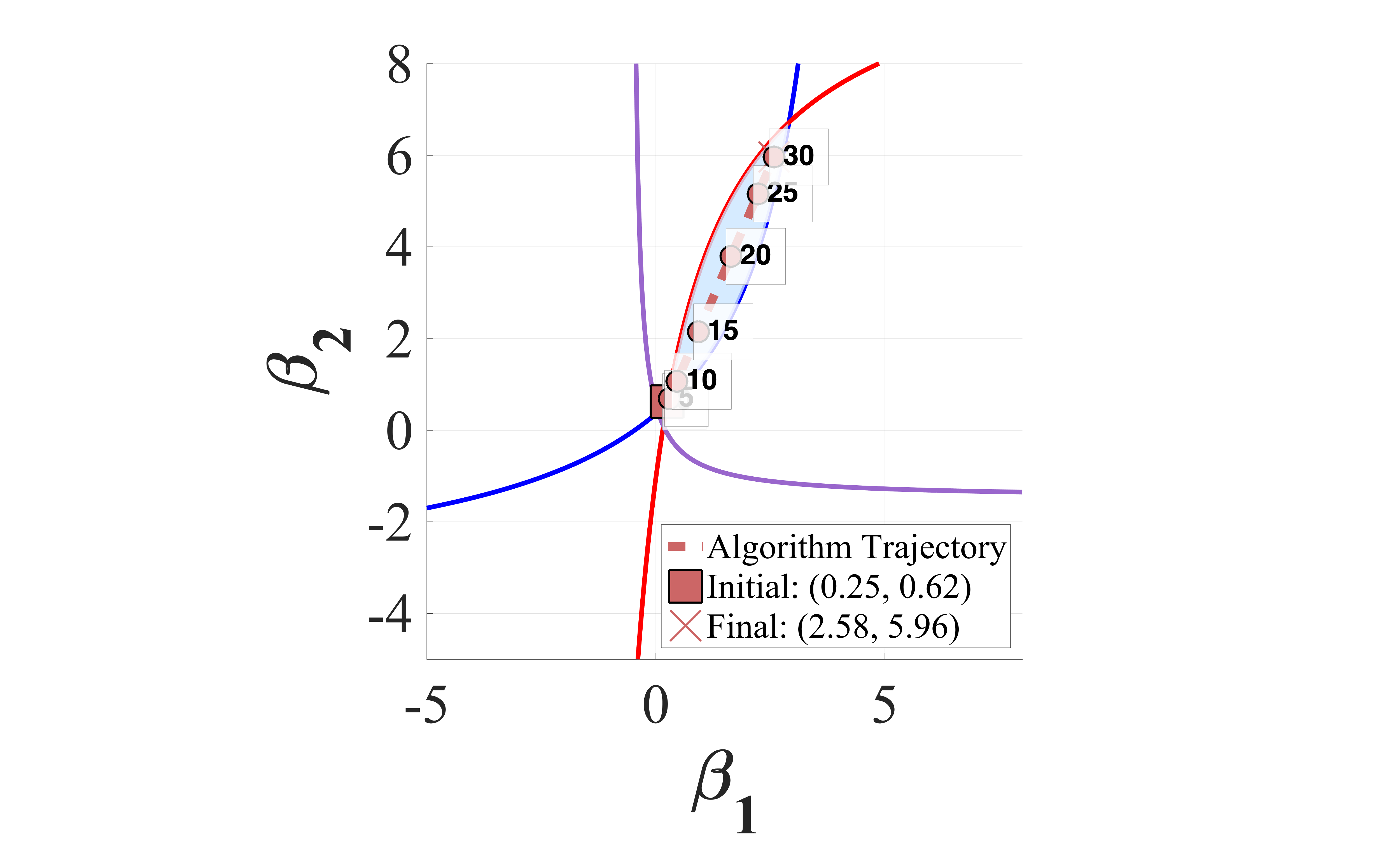} \\
(a) & (b) \\[0mm]
\includegraphics[width=0.18\textwidth, trim=210 0 300 0, clip]{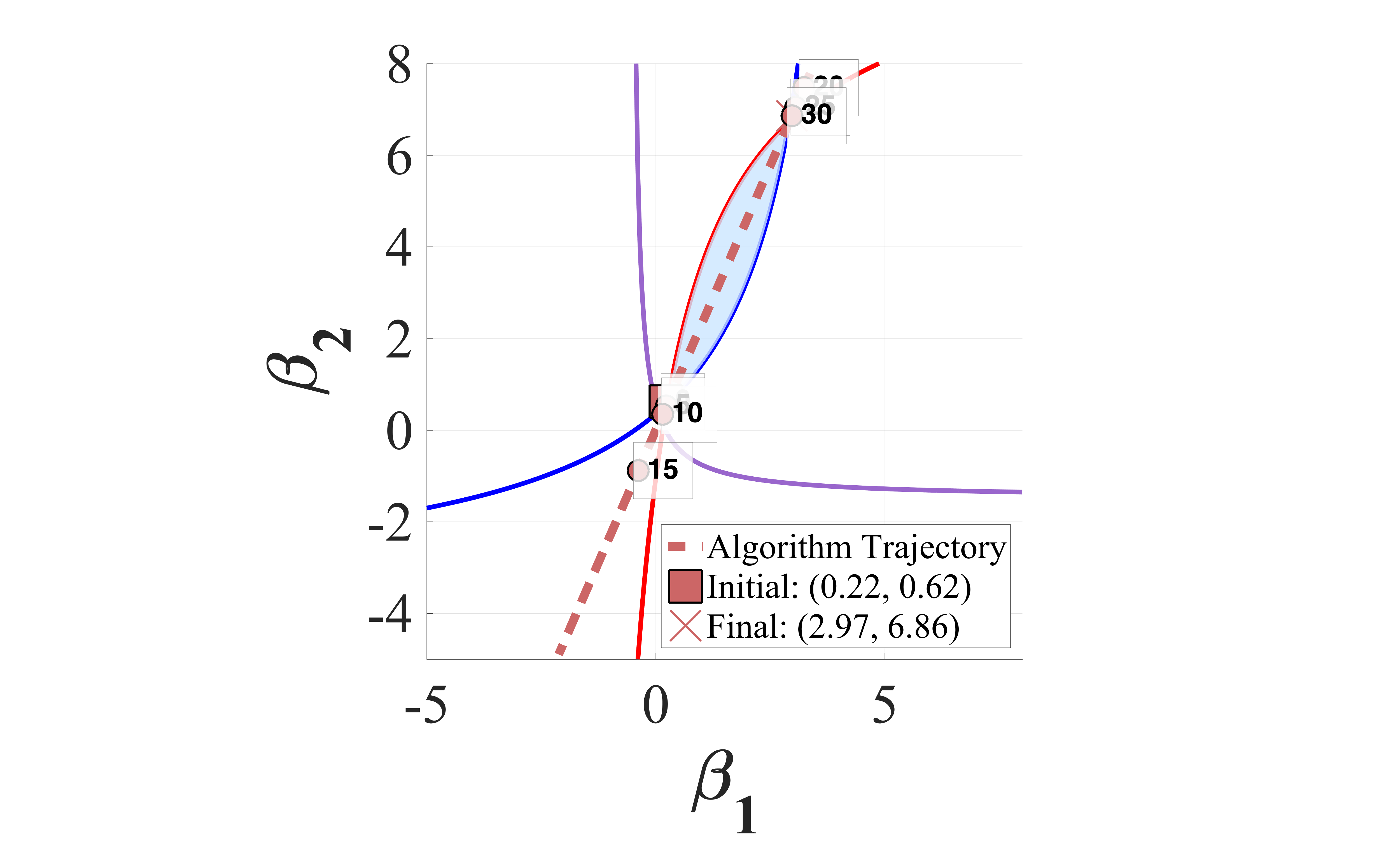} &
\includegraphics[width=0.18\textwidth, trim=190 0 300 0, clip]{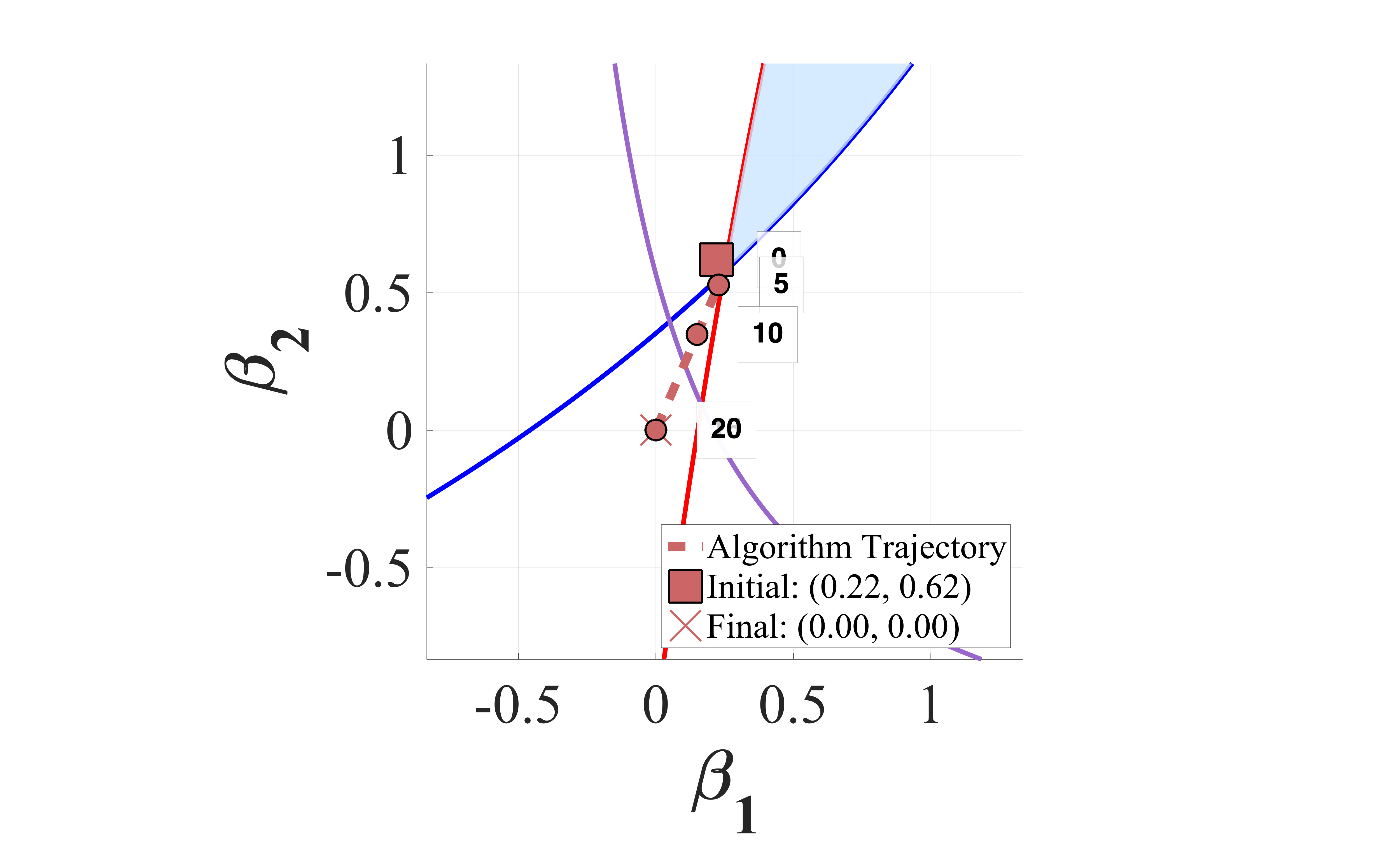} \\
(c) & (d)
\end{tabular}
\caption{Convergence behavior of the FPI \eqref{eq:fp_iter} using four different initialization points.}
\label{fig:alg_behaviour}
\vspace{-16pt}
\end{figure}

\begin{table}[t]
\centering
\caption{Comparison of Initialization Methods ($K=2$, $M=2$, $N=200$).}
\label{tab:init_comparison}
\begin{tabular}{lccc}
\toprule
Method & Outer Iter. & Inner Iter. & Init. Time (s) \\
\midrule
Fixed($0$)   & \textbf{17.4} & 277.7 & 0.0002 \\
Fixed($10$)  & 17.5 & 308.6 & \textbf{0.0001} \\
SDP          & 17.5 & \textbf{221.2} & 6.5068 \\
Warm-Start        & 17.5 & 250.9 & \textbf{0.0001} \\
Heuristic    & \textbf{17.4} & 277.7 & 0.0036 \\
\bottomrule
\end{tabular}
\vspace{-10pt}
\end{table}

\begin{figure*}[!t]
\centering
\begin{tabular}{cc}
\includegraphics[width=0.36\textwidth, trim=12 0 45 5, clip]{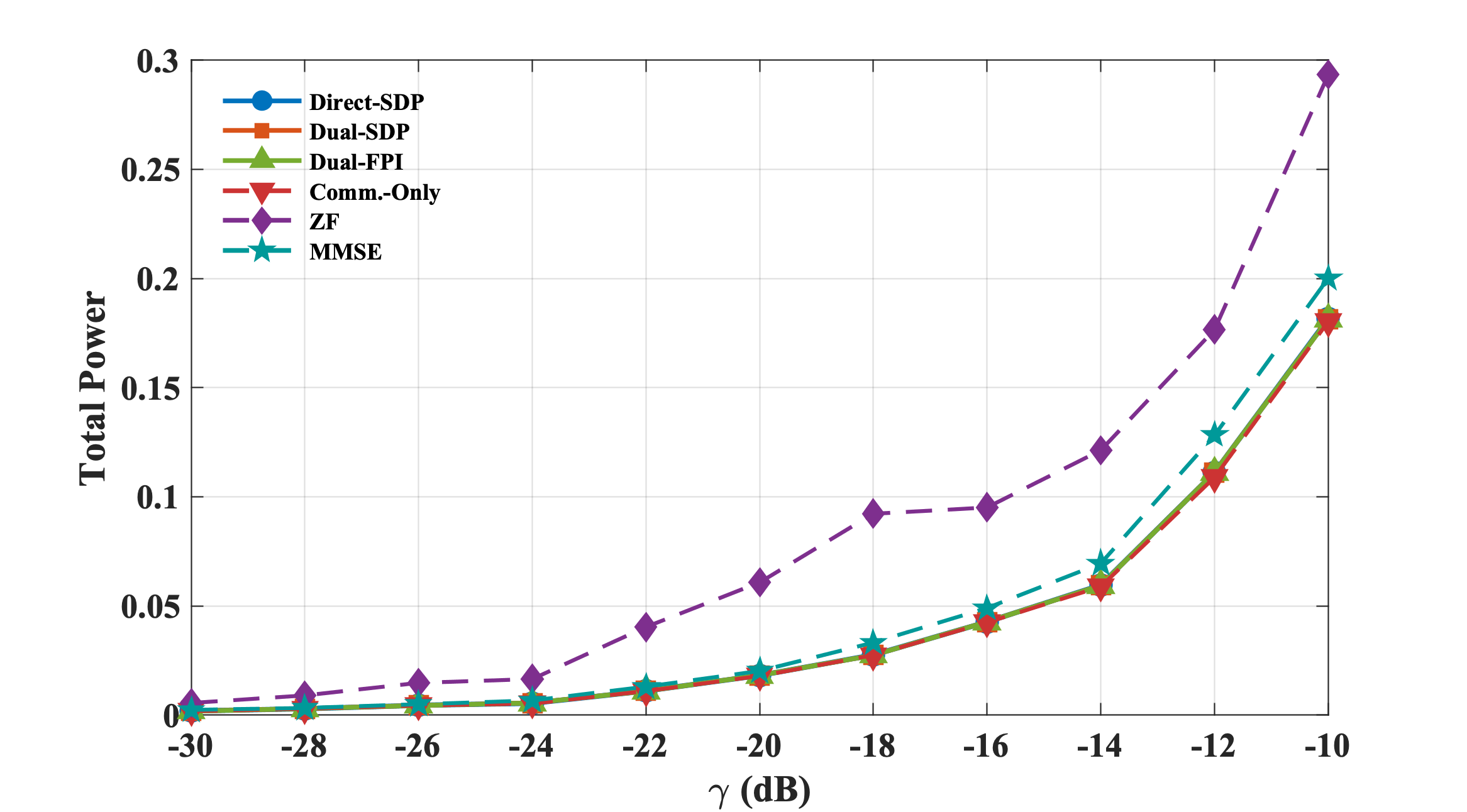} &
\includegraphics[width=0.36\textwidth, trim=15 0 45 5, clip]{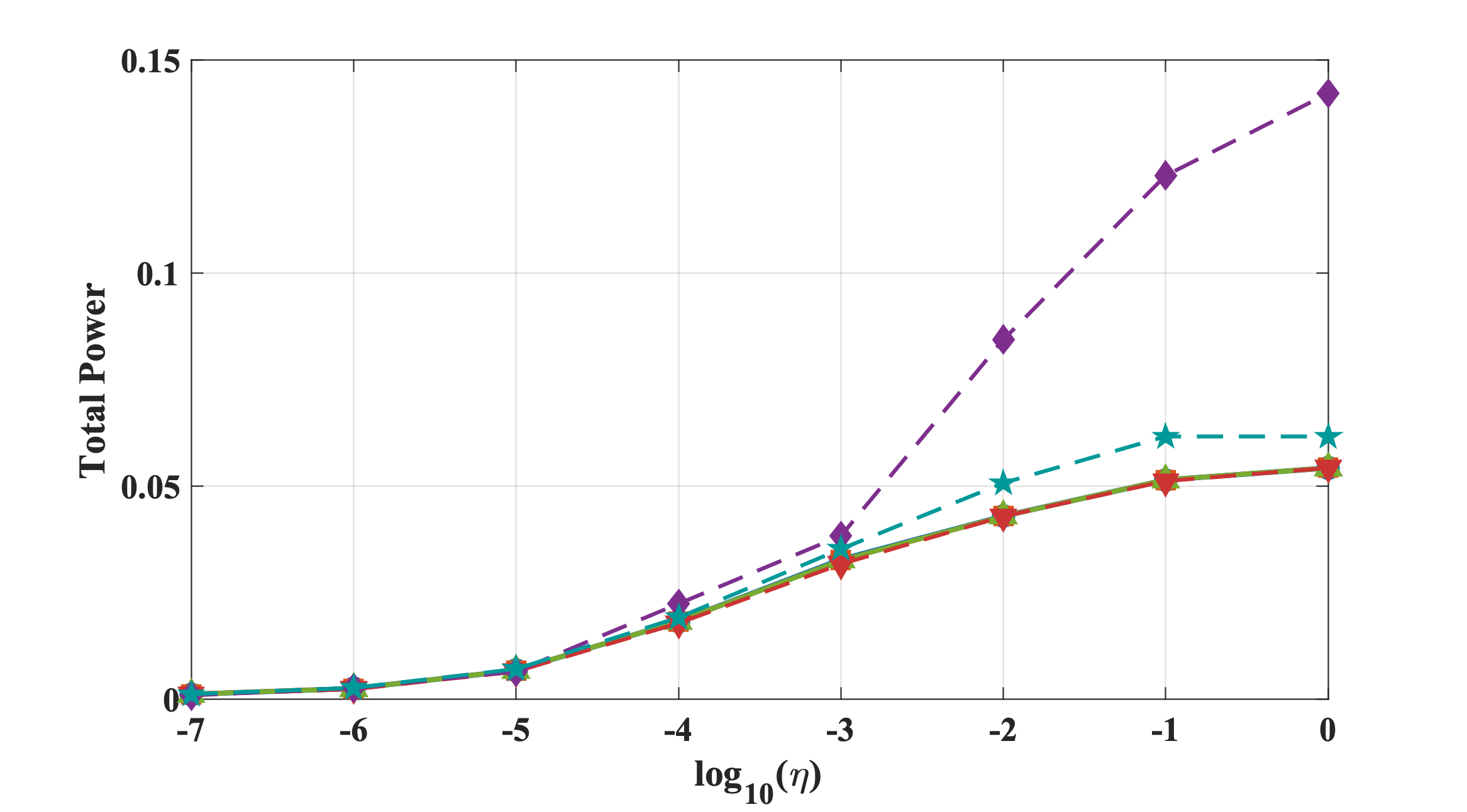} \\
(a) & (b) \\[1mm]
\includegraphics[width=0.36\textwidth, trim=15 0 45 5, clip]{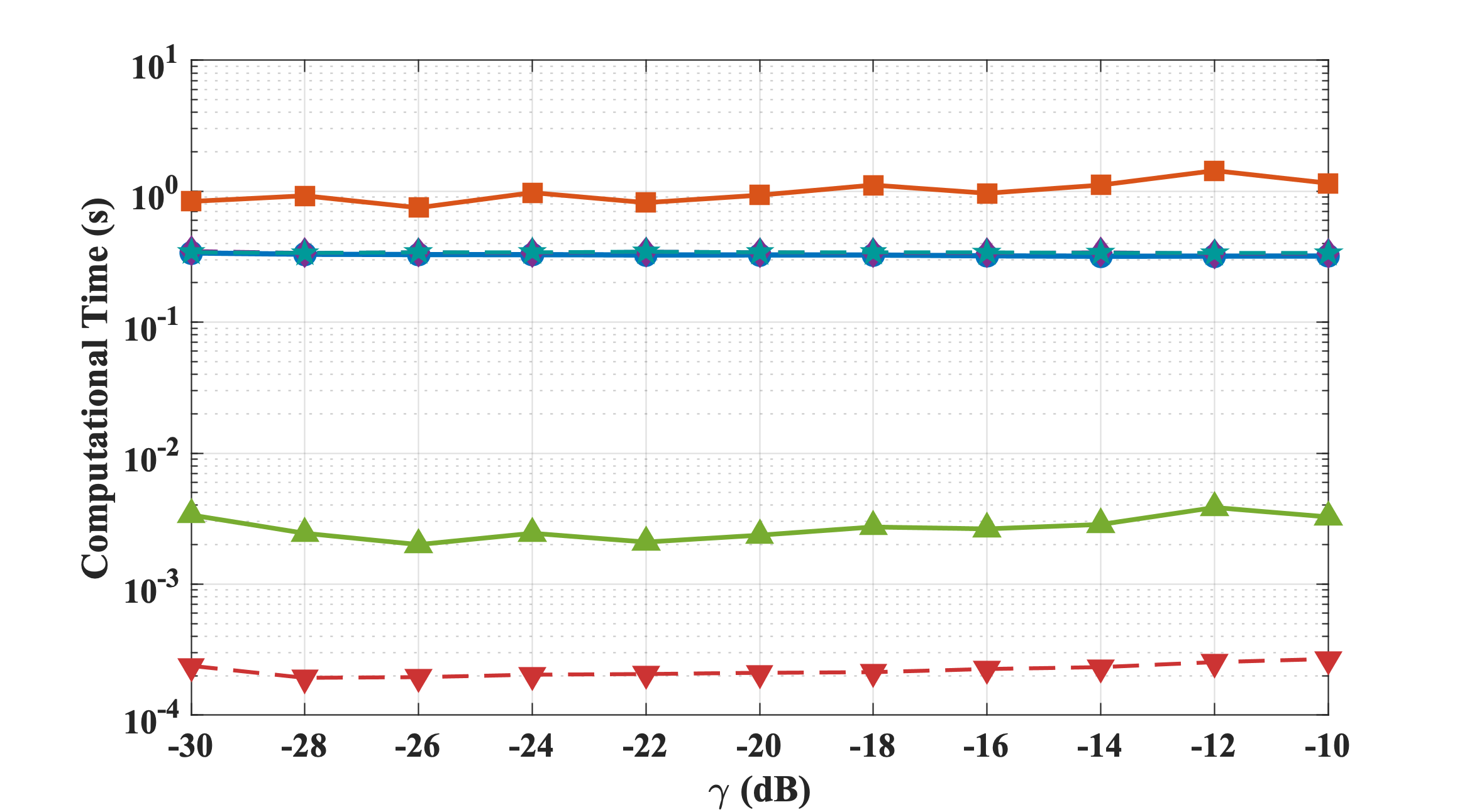} &
\includegraphics[width=0.36\textwidth, trim=15 0 45 5, clip]{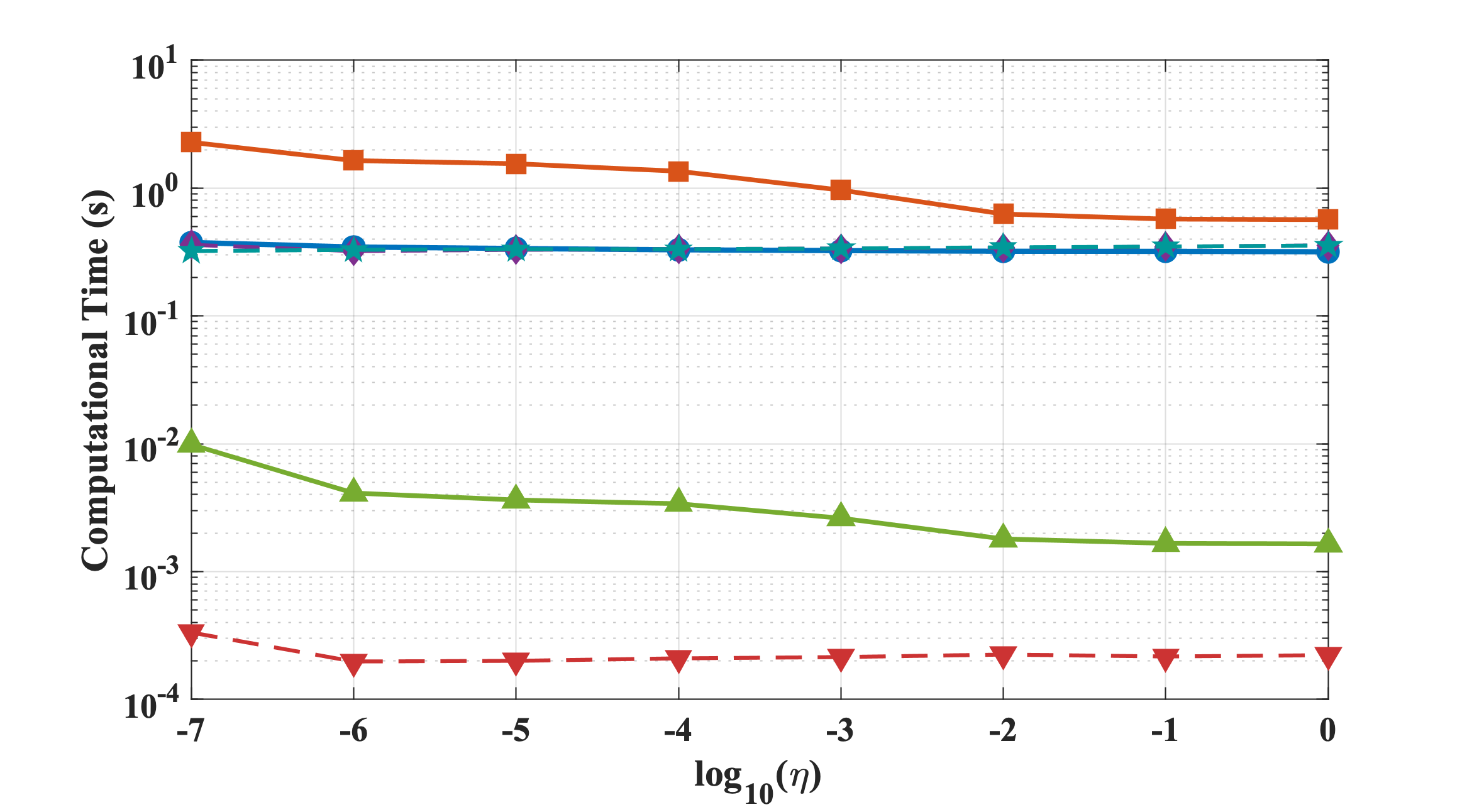} \\
(c) & (d)
\end{tabular}
\caption{Parameter sensitivity for $K = 2$ and $M = 2$. Top row: total transmit power. Bottom row: computational time. Left column: varying SINR threshold $\gamma$. Right column: varying MSE threshold~$\eta$.}
\label{fig:benchmark_single}
\vspace{-10pt}
\end{figure*}

To illustrate the importance of the initialization in the FPI algorithm in \eqref{eq:fp_iter}, we examine four initialization approaches: (a) strictly dual feasible initialization; (b) initialization within the convergent initialization region $\mathcal S$ in \eqref{eq:S} but dual infeasible; (c) initialization outside the convergent initialization region; and (d) the same initialization as (c), but using the FPI with a heuristic projection $\v \beta^{(i+1)} = \max \{I(\v \beta^{(i)}), \v 0\}$ to maintain nonnegativity.
Fig.~\ref{fig:alg_behaviour} presents the convergence trajectories of the FPI in \eqref{eq:fp_iter} for the GDB instance with $K = 2$ and $M = 2$ detailed in Appendix~\ref{apd:multiple_fp_example}.
The blue, red, and purple curves in the figure correspond to $\beta_1 = I_1(\v \beta)$, $\beta_2 = I_2(\v \beta)$, and $\det(\m C(\v \beta)) = 0$, respectively.
The feasible region is highlighted in light blue.

From Fig.~\ref{fig:alg_behaviour}, we observe the following.
(i) The feasible region contains two fixed points of $I(\cdot)$, but only the upper right one corresponds to the optimal dual solution as it maximizes the objective in problem \eqref{eq:dual2}.
This confirms the role of Proposition~\ref{prop:dual_characterization}~(a) in identifying the correct solution among multiple fixed points.
(ii) The FPIs initialized within the convergent region specified by Theorem~\ref{thm:convergence} (i.e., Figs.~\ref{fig:alg_behaviour}~(a) and~(b)) converge reliably to the optimal fixed point, validating the theoretical guarantees.
(iii) Although the initialization in Fig.~\ref{fig:alg_behaviour}~(c) is close to that of Fig.~\ref{fig:alg_behaviour}~(b), the iterates initially move toward the lower left before jumping to the upper right.
While this trajectory ultimately converges, the abrupt transition indicates potential numerical instability when the initialization lies outside $\mathcal S$.
(iv) The projection-based approach in Fig.~\ref{fig:alg_behaviour}~(d) fails to converge. The iterates become trapped at the origin, illustrating that the simple projection cannot substitute for a proper initialization within $\mathcal S$.

We next quantitatively compare the three initialization strategies from Section~\ref{sec:initialization} against two fixed-value baselines, denoted Fixed($0$) and Fixed($10$), which set $\beta_k^{(0)} = 0$ and $\beta_k^{(0)} = 10$ for all $k \in \cK$, respectively. 
Table~\ref{tab:init_comparison} reports the results averaged over $N = 200$ random instances with $K = 2$ and $M = 2$, where ``Outer Iter.'' denotes the average number of outer iterations, ``Inner Iter.'' denotes the average number of inner iterations in the FPI \eqref{eq:fp_iter}, and ``Init. Time'' denotes the average computational time to find the initial point in the FPI \eqref{eq:fp_iter}. We can make the following observations from Table \ref{tab:init_comparison}.
(i) All five methods require nearly the same number of outer iterations, indicating that the initialization strategy does not affect outer loop convergence.
(ii) The SDP-based Method~I yields the fewest inner iterations due to its strictly feasible starting point, but solving the SDP in \eqref{eq:sdp_slater} at each outer iteration incurs a cost orders of magnitude higher than any other method.
(iii) The warm-start Method~II achieves a comparable reduction in the number of inner iterations with negligible overhead, offering the best trade-off among all methods.
(iv) The heuristic Method~III performs similarly to the Fixed($0$) baseline in the inner iteration count, and Fixed($10$) results in the most inner iterations.

\subsection{Performance Comparison with SOTA Algorithms}

\begin{figure*}[!t]
\centering
\begin{tabular}{cc}
\includegraphics[width=0.36\textwidth, trim=15 0 45 5, clip]{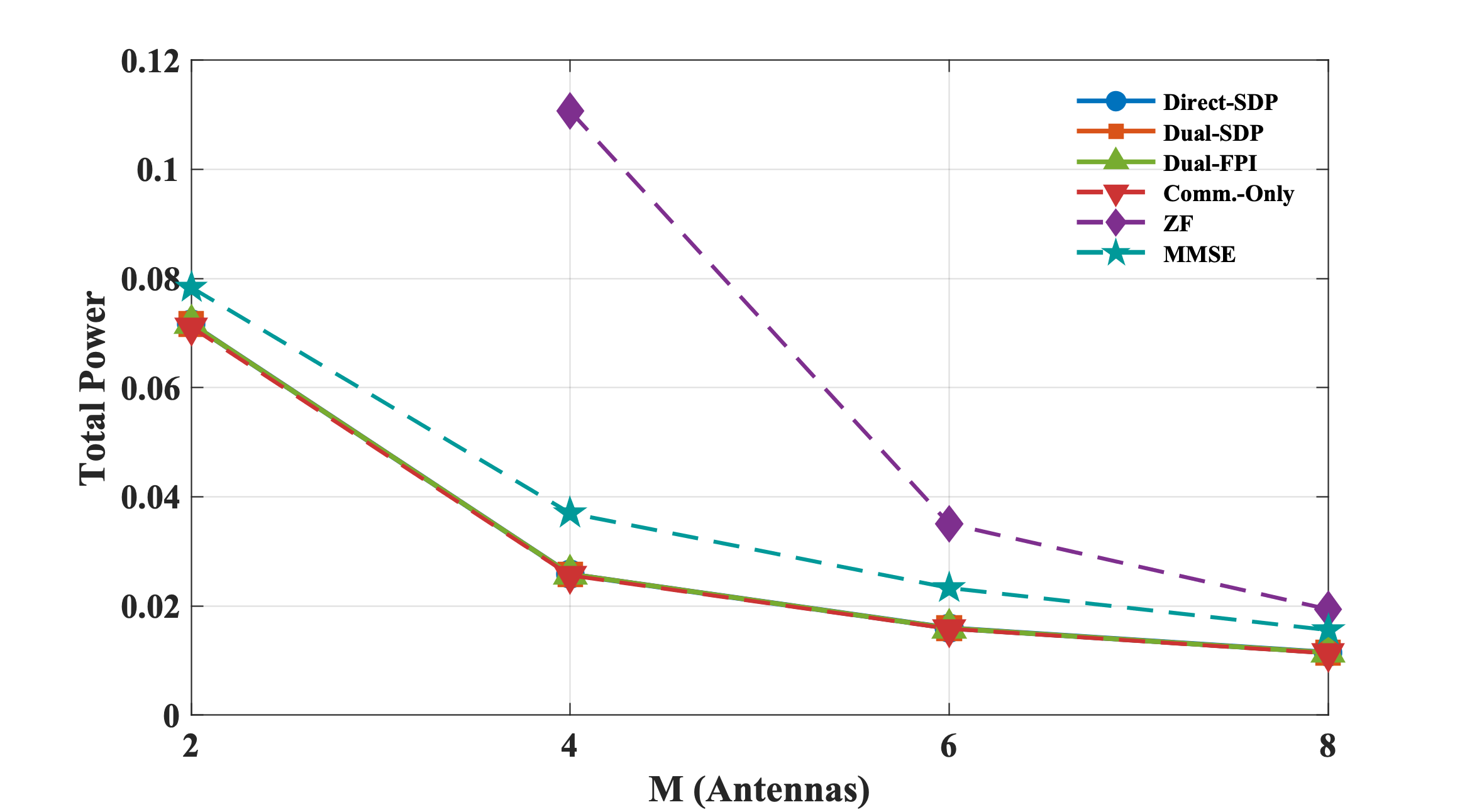} &
\includegraphics[width=0.36\textwidth, trim=15 0 45 5, clip]{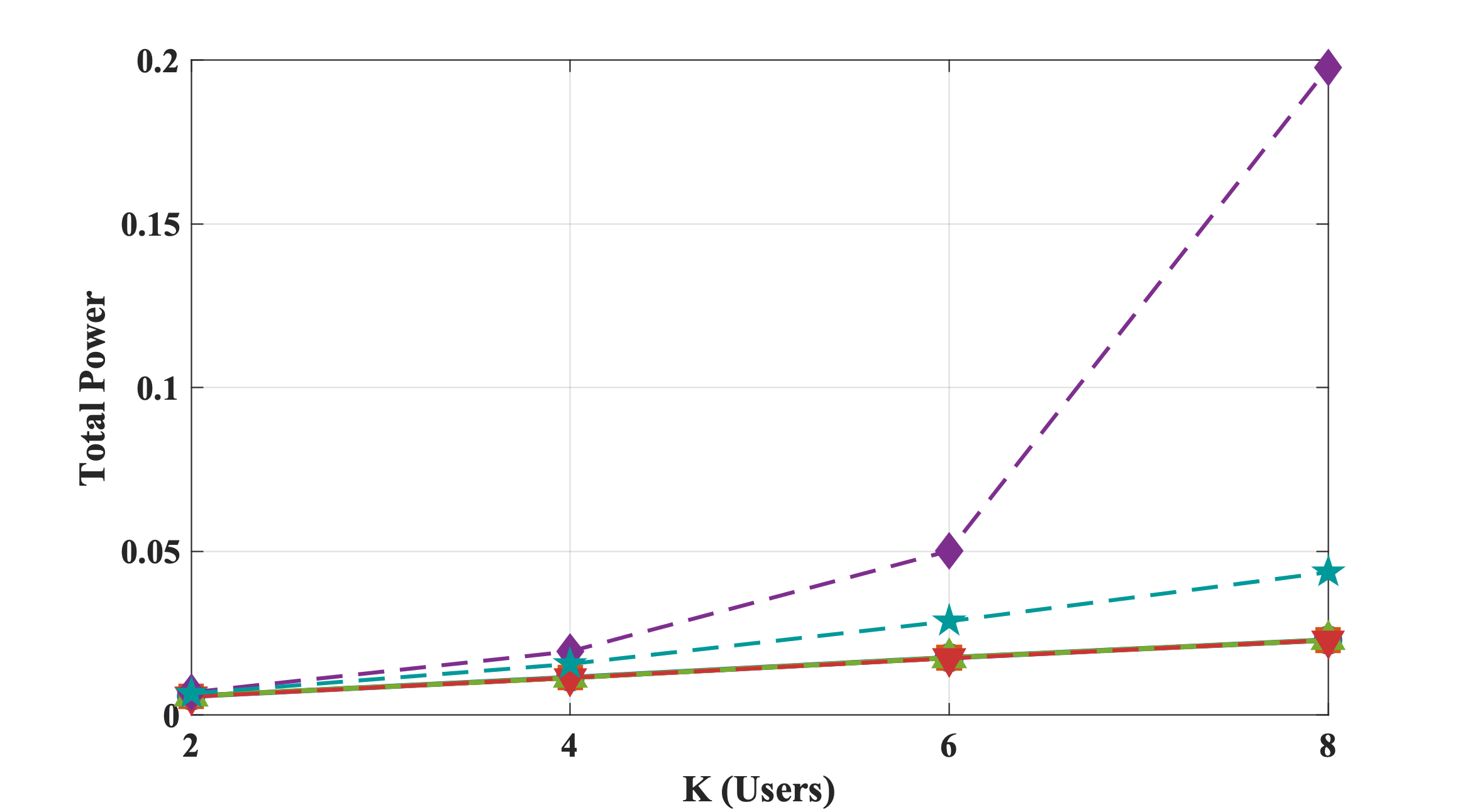} \\
(a) & (b) \\[1mm]
\includegraphics[width=0.36\textwidth, trim=15 0 45 5, clip]{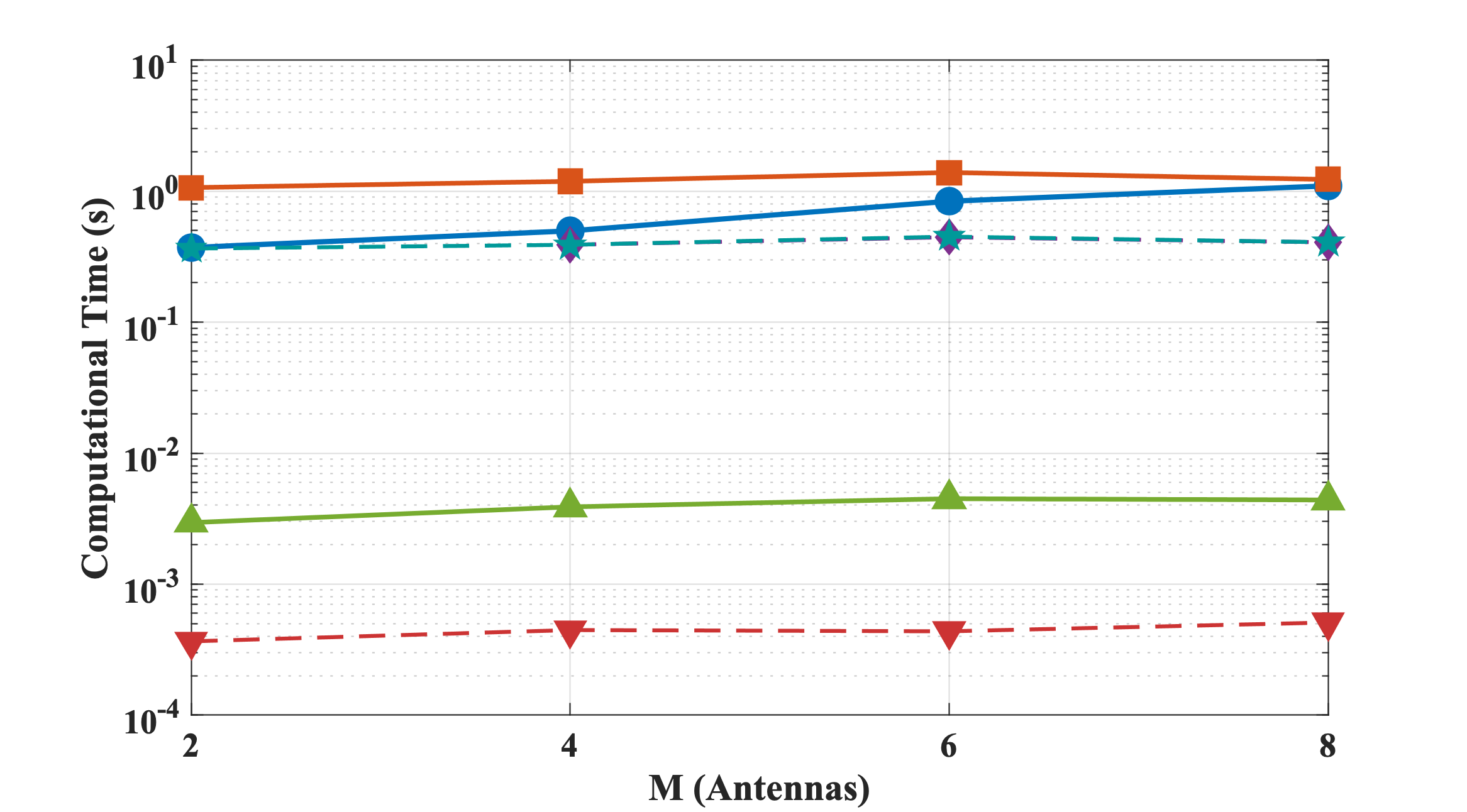} &
\includegraphics[width=0.36\textwidth, trim=15 0 45 5, clip]{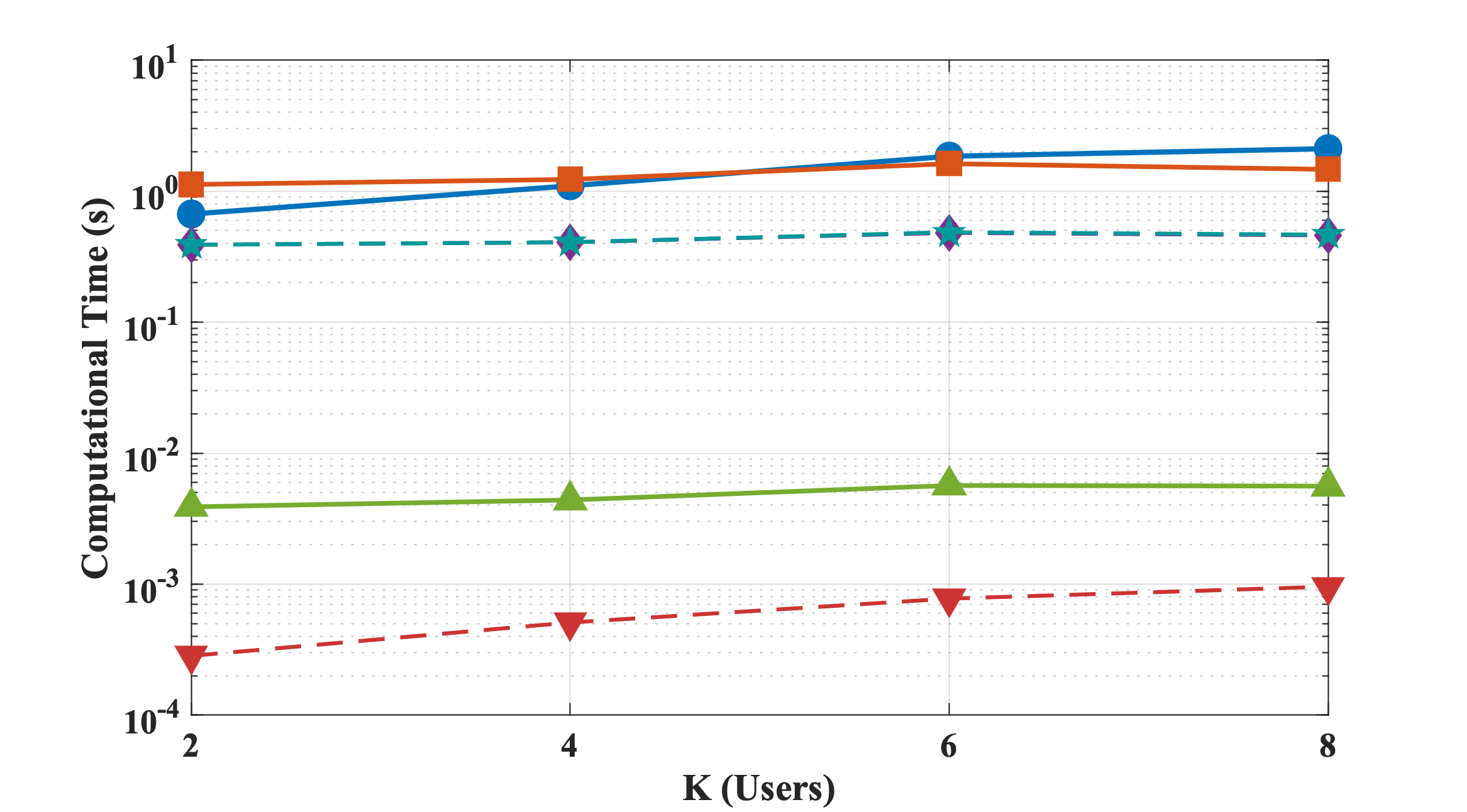} \\
(c) & (d)
\end{tabular}
\caption{Scalability comparison. Top row: total transmit power. Bottom row: computational time. Left column: varying $M$ with $K = 4$ fixed. Right column: varying $K$ with $M = 8$ fixed. ZF is not available when $K > M$.}
\label{fig:scalability}
\vspace{-8pt}
\end{figure*}

We compare the following six methods.
The first three solve the original ISAC problem \eqref{eq:sdp} using different algorithmic strategies:
\begin{itemize}
  \item \textbf{Direct-SDP:} the algorithm that solves problem \eqref{eq:sdp} directly using CVX with the SDPT3 solver.
  \item \textbf{Dual-SDP:} Algorithm~\ref{alg:dual_algorithm} with the inner GDB problem \eqref{eq:gdb} solved by the SDP solver.
  \item \textbf{Dual-FPI:} Algorithm~\ref{alg:dual_algorithm} with the inner GDB problem \eqref{eq:gdb} solved by Algorithm~\ref{alg:fp_algorithm}.
\end{itemize}
For the duality-based methods, we use the warm-start initialization from Section~\ref{sec:initialization}.
We also include the following three benchmarks for comparison:
\begin{itemize}
  \item \textbf{Comm.-Only:} the algorithm that solves the classical communication-only DB problem \cite{rashidDL1998} by dropping the sensing constraint, providing a power lower bound on the original ISAC problem.
  \item \textbf{ZF:} the algorithm that fixes the beamforming directions to zero-forcing vectors, then jointly optimizes the scalar powers and the sensing covariance matrix via SDP. This algorithm is only applicable when $K \leq M$.
  \item \textbf{MMSE:} the algorithm is the same as ZF but with MMSE beamforming directions.
\end{itemize}
The ZF and MMSE benchmarks adopt a suboptimal strategy inspired by \cite{zhang_joint_2025}.
The gap between the optimal objective and the Comm.-Only lower bound quantifies the power overhead due to the sensing constraint, while the gap between ZF/MMSE and the optimum reflects the benefit of jointly optimizing the beamforming directions. 
Other recent duality-based ISAC works \cite{attiah2025uplink-downlink,zhang_joint_2025,zhang2024optimal,zhu2024joint}, though closely related, cannot serve as direct benchmarks here due to fundamental differences in problem formulations.

\begin{table}[t]
  \centering
  \setlength{\tabcolsep}{4pt}
  \caption{Solution Quality Comparison for $K = 2$ and $M = 2$.}
  \label{tab:solution_quality_2_2}
  \begin{tabular}{cccc}
  \toprule
  Method & Obj Error & SINR Violation & MSE Violation \\ \midrule
  Direct-SDP & 0 & $0$ & $0$ \\
  Dual-SDP & $-1.04 \times 10^{-3}$ & $1.49 \times 10^{-16}$ & $8.69 \times 10^{-3}$ \\
  Dual-FPI & $-1.04 \times 10^{-3}$ & $1.73 \times 10^{-16}$ & $8.69 \times 10^{-3}$ \\
  Comm.-Only & $-8.64 \times 10^{-3}$ & $1.70 \times 10^{-16}$ & $1.55 \times 10^{-1}$ \\
  ZF & $1.95 \times 10^{0}$ & $1.95 \times 10^{-16}$ & $0$ \\
  MMSE & $2.27 \times 10^{-1}$ & $0$ & $0$ \\
  \bottomrule
  \end{tabular}
\end{table}

Table~\ref{tab:solution_quality_2_2} presents the solution quality for a representative instance with $K = 2$ and $M = 2$.
Direct-SDP, Dual-SDP, and Dual-FPI all achieve nearly identical objective values, with errors on the order of $10^{-3}$ relative to Direct-SDP, confirming global optimality.
Comm.-Only attains a lower objective as expected since the sensing constraint is dropped, and its MSE violation confirms that the power savings come at the cost of violating the sensing requirement.
Both~ZF and MMSE satisfy the constraints but incur significant suboptimality gaps, showing the benefit of jointly optimizing beamforming directions.

Fig.~\ref{fig:benchmark_single} shows how the six methods compare as the SINR threshold~$\gamma$ and the MSE threshold~$\eta$ vary for $K = 2$ and $M = 2$.
In subplots (a) and (b), Direct-SDP, Dual-SDP, and Dual-FPI produce overlapping objective curves, confirming global optimality across all parameter values.
ZF is consistently the worst, and MMSE lies between ZF and the optimal methods.
In subplots (c) and (d), Dual-FPI is about two orders of magnitude faster than Direct-SDP across all parameter values, while Dual-SDP is the slowest due to the repeated SDP solution at each outer iteration.

Fig.~\ref{fig:scalability} illustrates the scaling behavior as the number of antennas~$M$ and users~$K$ increase.
In subplots (a) and (b), Direct-SDP, Dual-SDP, and Dual-FPI achieve identical objective values across all tested configurations, while ZF and MMSE are consistently suboptimal with gaps that grow with the problem size.
ZF is not available when $K > M$, as zero-forcing requires at least as many antennas as users.
In subplots (c) and (d), the computational time of Dual-FPI remains nearly flat as $M$ and $K$ increase, whereas Direct-SDP grows notably.
As a result, Dual-FPI is about two orders of magnitude faster than Direct-SDP at smaller sizes, and the gap widens to nearly three orders of magnitude at larger sizes.
Overall, Dual-FPI attains globally optimal solutions with substantially lower computational cost than all competing approaches.

\section{Conclusion}\label{sec:conclusion}

In this paper, we investigate the multi-user ISAC beamforming design problem of minimizing the transmit power under communication SINR and radar MSE constraints.
The problem is reformulated as a GDB problem with possibly indefinite weighting matrices.
We characterize the necessary and sufficient condition for the boundedness of the GDB problem and propose an efficient FPI algorithm for solving it.
We further show that the maximal fixed point is the unique stable fixed point, while all other fixed points are unstable, and establish a linear convergence rate for the dual FPI.
We also propose three initialization strategies with different trade-offs between theoretical guarantees and computational costs.
Based on these results, we develop a Dual-FPI algorithm for solving the considered ISAC beamforming design problem and establish its overall convergence guarantee.
Simulations demonstrate that the proposed Dual-FPI algorithm achieves globally optimal solutions with substantially lower computational cost than existing approaches.

\appendices
\section{An Illustrative Instance with Two Fixed Points}\label{apd:multiple_fp_example}

We construct an explicit example in which the interference function $I(\cdot)$ in \eqref{eq:interference} admits two fixed points.
This is also the instance underlying Fig.~\ref{fig:alg_behaviour}.

Consider $K = 2$ users and $M = 2$ antennas with channels
\begin{equation}
\v h_1 = \begin{bmatrix}1.2 \\ 0.8\end{bmatrix},\quad
\v h_2 = \begin{bmatrix}0.9 \\ 1.1\end{bmatrix},
\label{eq:apd_channels}
\end{equation}
weighting matrix $\m B = \mathrm{diag}(-0.2,\, 1.0)$, SINR targets $\gamma_1 = \gamma_2 = 3$, and noise powers $\sigma_1^2 = \sigma_2^2 = 1$.
The matrix $\m B$ has eigenvalues $-0.2$ and $1.0$ and is therefore indefinite.

\begin{figure}[t]
\centering
\includegraphics[width=0.3\textwidth]{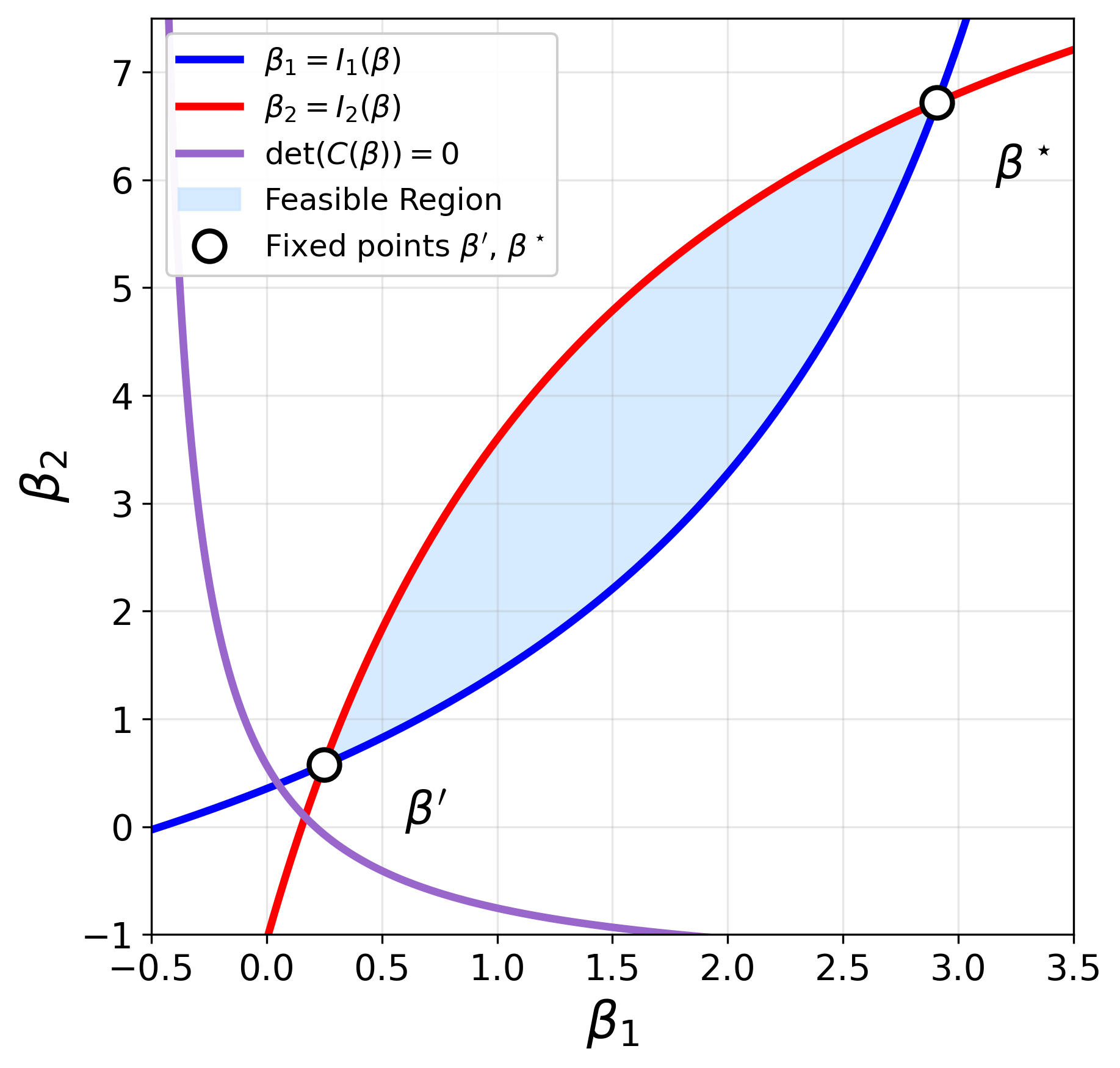}
\caption{Visualization of the example in Appendix~\ref{apd:multiple_fp_example}.
The curve and region conventions follow Fig.~\ref{fig:alg_behaviour}.
The two intersections of the blue and red curves are the fixed points $\v \beta'$ and $\v \beta^\star$.}
\label{fig:apd_two_fps}
\vspace{-10pt}
\end{figure}

Substituting $\m C(\v \beta)^{-1}$ into \eqref{eq:interference} and clearing denominators reduces the fixed point equation $\v \beta = I(\v \beta)$ to a quadratic equation in $\beta_1$, which has two real positive roots.
The corresponding fixed points are
\begin{equation*}
\v \beta' \approx (0.2482,\ 0.5733),\quad \v \beta^\star \approx (2.9074,\ 6.7156),
\end{equation*}
as visualized in Fig.~\ref{fig:apd_two_fps}.
Both $\m C(\v \beta')$ and $\m C(\v \beta^\star)$ are positive definite, so $\v \beta', \v \beta^\star \in \mathcal F_1$.
Since $\v \beta^\star > \v \beta'$ component-wise, $\v \beta^\star$ is the maximal fixed point and coincides with the optimal dual solution by Proposition~\ref{prop:dual_characterization}~(a).

\section{Proof of Theorem~\ref{thm:fpi_convergence_rate}}\label{apd:fpi_convergence_rate}

We prove Theorem~\ref{thm:fpi_convergence_rate} by establishing a contraction property of the dual FPI under the metric $\mu(\cdot, \cdot)$ defined in \eqref{eq:metric}.
Recall that the dual FPI can be expressed in terms of the shifted variable $\v{\delta}^{(i)} = \v{\beta}^{(i)} - \hat{\v{\beta}}$ as $\v{\delta}^{(i+1)} = J(\v{\delta}^{(i)})$, where $J(\v{\delta}) = I(\hat{\v{\beta}} + \v{\delta}) - \hat{\v{\beta}}$ with fixed point $\v{\delta}^\star = \v{\beta}^\star - \hat{\v{\beta}}$.

The proof consists of three key steps.
First, we establish an upper bound on $\mu(J(\v y), J(\v z))$ in terms of $\mu(\v y, \v z)$.
Second, we bound the ratio ${J(\alpha \v \delta)}/{J(\v \delta)}$ using the spectral properties of the shifted interference function.
Finally, we combine these results to obtain the convergence rate.

\textit{Step 1: Upper bound on the metric:}
For any $\v y, \v z \in \mathbb{R}_{++}^K$, let $\alpha = \mathrm{e}^{\mu(\v y, \v z)}$.
By the definition of $\mu(\cdot, \cdot)$, we have $\v y \leq \alpha \v z$ and $\v z \leq \alpha \v y$.
Combining this with the monotonicity of $I(\cdot)$ gives
\begin{equation}
I(\hat{\v \beta} + \v y) \leq I(\hat{\v \beta} + \alpha \v z) \quad \text{and} \quad I(\hat{\v \beta} + \v z) \leq I(\hat{\v \beta} + \alpha \v y).
\end{equation}
Since $J(\v \delta) = I(\hat{\v \beta} + \v \delta) - \hat{\v \beta}$, we have
\begin{equation}
J(\v y) \leq J(\alpha \v z) \quad \text{and} \quad J(\v z) \leq J(\alpha \v y).
\end{equation}
As a result,
\begin{equation}
\mu(J(\v y), J(\v z)) \leq \max_k \left\{ \log \left( \frac{J_k(\alpha \v z)}{J_k(\v z)} \right), \log \left( \frac{J_k(\alpha \v y)}{J_k(\v y)} \right) \right\}.
\label{eq:mu_bound_J}
\end{equation}

\textit{Step 2: Bound on the ratio:}
For any $\alpha > 1$ and $\v \delta \in \mathbb{R}_{++}^K$, we have
\begin{equation}
\frac{J(\alpha \v \delta)}{J(\v \delta)} = \frac{I(\hat{\v \beta} + \alpha \v \delta) - \hat{\v \beta}}{I(\hat{\v \beta} + \v \delta) - \hat{\v \beta}}.
\end{equation}
Let $a = I(\hat{\v \beta} + \alpha \v \delta)$, $b = I(\hat{\v \beta} + \v \delta)$, and $x = \hat{\v \beta}$.
Since $\hat{\v \beta} < I(\hat{\v \beta})$ by Slater's condition and $\alpha > 1$, we have $0 \leq x < b < a$.
Therefore,
\begin{equation}
\frac{a - x}{b - x} \leq \frac{a}{b},
\end{equation}
which gives
\begin{equation}
\frac{J(\alpha \v \delta)}{J(\v \delta)} \leq \frac{I(\hat{\v \beta} + \alpha \v \delta)}{I(\hat{\v \beta} + \v \delta)}.
\label{eq:J_ratio_ineq}
\end{equation}

Now, by the definition of $I_k(\cdot)$ in \eqref{eq:interference}, we have
\begin{equation}
I_k(\hat{\v \beta} + \v \delta) = \frac{\gamma_k}{\gamma_k + 1} \frac{1}{\v h_k^\hermitian \left( \m B + \sum_{j \in \cK} (\hat{\beta}_j + \delta_j) \v h_j \v h_j^\hermitian \right)^{-1} \v h_k}.
\end{equation}
Define $\hat{\m B} = \m B + \sum_{j \in \cK} \hat{\beta}_j \v h_j \v h_j^\hermitian$, which is positive definite by Slater's condition.
Then
\begin{equation}
I_k(\hat{\v \beta} + \v \delta) = \frac{\gamma_k}{\gamma_k + 1} \frac{1}{\v h_k^\hermitian \left( \hat{\m B} + \sum_{j \in \cK} \delta_j \v h_j \v h_j^\hermitian \right)^{-1} \v h_k},
\end{equation}
and similarly
\begin{equation}
I_k(\hat{\v \beta} + \alpha \v \delta) = \frac{\gamma_k}{\gamma_k + 1} \frac{1}{\v h_k^\hermitian \left( \hat{\m B} + \sum_{j \in \cK} \alpha \delta_j \v h_j \v h_j^\hermitian \right)^{-1} \v h_k}.
\end{equation}
Combining the above with \eqref{eq:J_ratio_ineq} yields, for each $k \in \cK$,
\begin{equation}
\frac{I_k(\hat{\v \beta} + \alpha \v \delta)}{I_k(\hat{\v \beta} + \v \delta)} = \frac{\v h_k^\hermitian \left( \hat{\m B} + \sum_{j \in \cK} \delta_j \v h_j \v h_j^\hermitian \right)^{-1} \v h_k}{\v h_k^\hermitian \left( \hat{\m B} + \sum_{j \in \cK} \alpha \delta_j \v h_j \v h_j^\hermitian \right)^{-1} \v h_k}.
\end{equation}

Define $\lambda(\v \delta) = \rho(\hat{\m B} \sum_{j \in \cK} \delta_j \v h_j \v h_j^\hermitian)$.
Using the eigenvalue interlacing property \cite[Theorem~A.1]{nocedal2006NumericalOptimization} and Weyl's inequality \cite[Theorem~4.3.1]{horn2012MatrixAnalysis}, it can be shown that
\[
\frac{\v h_k^\hermitian \left( \hat{\m B} + \sum_{j \in \cK} \delta_j \v h_j \v h_j^\hermitian \right)^{-1} \v h_k}{\v h_k^\hermitian \left( \hat{\m B} + \sum_{j \in \cK} \alpha \delta_j \v h_j \v h_j^\hermitian \right)^{-1} \v h_k}
\leq \frac{1 + \alpha \lambda(\v \delta)}{1 + \lambda(\v \delta)}.
\]
Taking the $\alpha$-logarithm gives
\begin{equation}
\log_\alpha \left( \frac{J(\alpha \v \delta)}{J(\v \delta)} \right) \leq \log_\alpha \left( \frac{1 + \alpha \lambda(\v \delta)}{1 + \lambda(\v \delta)} \right).
\label{eq:ratio_bound_J}
\end{equation}

\textit{Step 3: Convergence rate:}
Define $\kappa(\alpha, \lambda) = \log_\alpha \left( \frac{1+\alpha \lambda}{1+\lambda} \right)$.
It can be verified that $\kappa(\alpha, \lambda) \in (0,1)$ for any $\alpha > 1$ and $\lambda > 0$, and $\lim_{\alpha \rightarrow 1^+} \kappa(\alpha, \lambda) = {\lambda}/{(1+\lambda)}$.
From \eqref{eq:mu_bound_J} and \eqref{eq:ratio_bound_J}, we have
\[
\frac{\mu(\v{\delta}^{(i+1)}, \v{\delta}^\star)}{\mu(\v{\delta}^{(i)}, \v{\delta}^\star)} \leq \max \left\{ \kappa(\alpha_i, \lambda(\v{\delta}^{(i)})), \kappa(\alpha_i, \lambda(\v{\delta}^\star)) \right\},
\]
where $\alpha_i = \mathrm{e}^{\mu(\v{\delta}^{(i)}, \v{\delta}^\star)}$ and $\lambda(\v \delta) = \rho(\hat{\m B} \sum_{j \in \cK} \delta_j \v h_j \v h_j^\hermitian)$.
Taking the limit superior on both sides and noting that $\lim_{i \to \infty} \alpha_i = 1$ and $\lim_{i \to \infty} \lambda(\v{\delta}^{(i)}) = \lambda(\v{\delta}^\star)$, we obtain
\[
\limsup_{i\rightarrow \infty}\frac{\mu(\v{\delta}^{(i+1)}, \v{\delta}^\star)}{\mu(\v{\delta}^{(i)}, \v{\delta}^\star)} \leq \lim_{i \rightarrow \infty} \kappa(\alpha_i, \lambda(\v{\delta}^\star)) = \frac{\lambda(\v{\delta}^\star)}{1 + \lambda(\v{\delta}^\star)}.
\]
Since $\lambda(\v{\delta}^\star) = \rho(\hat{\m B} \sum_{j \in \cK} \delta_j^\star \v h_j \v h_j^\hermitian) = \lambda$ by definition, we have
\[
\frac{\lambda(\v{\delta}^\star)}{1 + \lambda(\v{\delta}^\star)} = \frac{\lambda}{1 + \lambda}.
\]
This completes the proof of \eqref{eq:c_r_limit}.
\hfill $\square$

\bibliographystyle{IEEEtran}
\bibliography{IEEEabrv, all_references}

\end{document}